\newif\iffigures
\figurestrue
\documentclass[aip,jmp,amsmath,amssymb,reprint]{revtex4-1}
\usepackage{amsmath}
\usepackage{xspace}%fix spacing after newcommands
\usepackage{dcolumn}%align table columns on decimal point
\usepackage{bm}%bold math
\usepackage{graphicx}

\usepackage[normalem]{ulem}
\usepackage{array}
\newcolumntype{C}[1]{>{\centering\let\newline\\\arraybackslash\hspace{0pt}}m{#1}}
\newcommand{\ai}{\emph{ab initio}\xspace}
\newcommand{\indn}{\ensuremath{V_{ind}^{NB}}\xspace}
\newcommand{\indthree}{\ensuremath{V_{ind}^{3B}}\xspace}
\newcommand{\indtwo}{\ensuremath{V_{ind}^{2B}}\xspace}
\newcommand{\polythree}{\ensuremath{V_{poly}^{3B}}\xspace}
\newcommand{\electwo}{\ensuremath{V_{elec}^{2B}}\xspace}
\newcommand{\hbbtwo}{\ensuremath{V_{HBB2}^{2B}}\xspace}
\newcommand{\ps}{\ensuremath{V_{PS}^{1B}}\xspace}
\begin{document}
\title[Two-body and three-body interactions in water]
{A critical assessment of two-body and three-body interactions in water}

\author{Gregory R. Medders}
\altaffiliation{Contributed equally to this work}
\author{Volodymyr Babin}
\altaffiliation{Contributed equally to this work}
\author{Francesco Paesani}
\email{fpaesani@ucsd.edu}
\affiliation
{Department of Chemistry and Biochemistry, University of California, San Diego, La Jolla, CA 92103}

\date{\today}

\begin{abstract}
The microscopic behavior of water under different conditions and
in different environments remains the subject of intense debate.
A great number of the controversies arises due to the contradictory
predictions obtained within different theoretical models.
Relative to conclusions derived
from force fields or density functional theory, there is comparably
less room to dispute highly-correlated electronic structure calculations.
Unfortunately, such \ai calculations are severely limited by system size.
In this study, a detailed analysis of the two- and three-body water
interactions evaluated at the CCSD(T) level is carried out to
quantitatively assess the accuracy of several force fields, density
functional theory, and \ai-based interaction potentials that are
commonly used in molecular simulations. Based on this analysis, a new
model, HBB2-pol, is introduced which is capable of accurately mapping
CCSD(T) results for water dimers and trimers into an efficient analytical
function. The accuracy of HBB2-pol is further established through
comparison with the experimentally determined second and third virial
coefficients.
\end{abstract}

\maketitle

\section{\label{sec:1} Introduction}

Connecting small clusters of water and the condensed phases of water through a single
molecular model has been a long sought-after but so-far unachieved goal. The challenges involved 
in the pursuit of this goal are numerous. 
For example, at the cluster level, the Born-Oppenheimer energies of topologically distinct isomers of the water
hexamer differ by less than 1 kcal/mol
\cite{Dahlke2008,Bates2009a,Gora2011}, indicating that highly-correlated electronic
structure calculations are required to quantitatively determine the energy order of these isomers.  
In this regard, a faithful description of molecular flexibility appears to be particularly important.\cite{Gora2011}
Furthermore, it has also been shown that nuclear quantum-mechanical effects 
 can impact the structural, thermodynamic, and dynamical 
properties of both clusters and bulk phases of water.\cite{Soper2008,Paesani2009,Wang2012b,Markland2012}
 The explicit inclusion of nuclear quantum effects in simulations exacerbates the
computational expense of a model, providing additional strain on the ability to obtain statistically
meaningful results.

The majority of water simulations rely on force fields, 
which are built upon the many-body expansion of the interaction energy,\cite{Hankins1970}
\begin{align} \label{eq:many-body}
  E&(1,\dots,N) = \sum_i^N V^{1B}(i) + \sum_{i<j}^N V^{2B}(i,j)\nonumber\\ 
                &+ \sum_{i<j<k}^N V^{3B}(i,j,k) +\dots + V^{NB}(1,\dots,N).
\end{align}
Here, $V^{1B}(i) = E(i) - E_{eq}(i)$ is the one-body (1B) potential, which describes the energy required 
to deform an individual molecule from its equilibrium geometry.
In common force-fields, the 1B interactions include all bonded terms (i.e., stretches, bends, and torsions). 
For systems, such as water, that are easily reduced to distinct molecules, a 1B configuration
is typically referred to as a ``monomer'', and groups of 2, 3, ... , $N$
interacting monomers are then termed ``dimers'', 
``trimers'', ..., ``$N$-mers''.
In Eq.~\ref{eq:many-body}, higher-order interactions are defined recursively through the 
lower-order terms. For instance, 
the two-body (2B) interaction is expressed as 
\begin{align}
  V^{2B} = E(1,2) - \sum_{i=1}^2 E(i)\nonumber
\end{align}
where $E(1,2)$ is the dimer energy. Similarly, the three-body (3B) interaction is
\begin{align}
  V^{3B} = E(1,2,3) - \sum_{i<j}^3 E(i,j) + \sum_{i=1}^3 E(i)\nonumber
\end{align}
with $E(1,2,3)$ being the trimer energy.
Common force fields are pairwise additive, 
 meaning that three-body and higher interactions are neglected.

If it converges quickly, the many-body expansion represents a
powerful approach to studying condensed phases as it allows for the energy of
an $N$-molecule system to be expressed as a sum of lower-order interactions
that can in principle be calculated with high accuracy.
Recently, a detailed study of the convergence of the many-body expansion for water
based on the analysis of small clusters was performed using 
coupled cluster theory with single, double, and perturbative triple excitations [CCSD(T)] and
large basis sets.\cite{Gora2011} 
Consistent with previous observations \cite{Xantheas1994,Xantheas2000,Defusco2007,%
Kumar2010,Hodges1997,Ojamie1994,Pedulla1996,Cui2006}, it was determined that, 
although two-body interactions dominate the expansion,
the three-body term can contribute up to 30\% of the total energy of the water hexamer. 
An estimate of the relative magnitudes of the many-body terms in 
liquid was obtained through an RIMP2 analysis of the 21-mer, for which two-body interactions were found to
contribute 75-80\% of the total interaction energy and three-body interactions comprised 
15-20\%.\cite{Cui2006} For both the water hexamer and the 21-mer, higher-order terms contribute 
less than 5\% of the total interaction energy. 
It should be noted that, while quickly converging for water, 
 the many-body expansion has been shown to 
converge slowly and with marked oscillatory behavior for other systems.\cite{Hermann2007}

In this study, the accuracy of several force fields, density functional theory (DFT), and \ai potentials
 in reproducing the two- and three-body water interactions is assessed through a
detailed comparison with data obtained at the CCSD(T) level of theory (Section \ref{sec:2}). 
Based on this analysis, a new \ai water model, HBB2-pol, is then introduced in Section \ref{sec:methods}. 
In Section \ref{sec:results}, we show that HBB2-pol accurately maps the CCSD(T) results for 
both the 2B and 3B interactions into an efficient analytical functional form and predicts the second
and third virial coefficients in excellent agreement with the available experimental data. 
A summary is given in Section \ref{sec:summary}.

\section{\label{sec:2} Analysis of two- and three-body water interactions}
\subsection{\label{sec:2-1}Water models}

The many-body expansion provides the underlying basis for common classical force fields.
In most cases, including the widely-used
TIP4P and SPC families\cite{Jorgensen1983,Berendsen1981},
pairwise additivity is assumed, with three- and higher-body interactions
being ``encoded" into the effective two-body contributions. In addition,
the majority of these models treat the water molecules as rigid monomers
(i.e., the 1B interactions are set to zero), with only few quantum water models, notably as
q-TIP4P/f and qSPC/Fw,\cite{Habershon2009,Paesani2006} allowing for
molecular flexibility. Nevertheless, pairwise force fields have
been surprisingly successful at reproducing, at least qualitatively,
the properties of water in homogeneous environments.\cite{Vega2011} 
However, such force fields are expected to be inherently limited in their 
ability to model the microscopic behavior of aqueous interfaces, water
confined at the nanoscale, and clusters, whose properties are
sensitive to the detailed interplay of 1B, 2B, 3B, and
higher-body interactions.\cite{Xantheas1994,Xantheas2000,Defusco2007,Kumar2010,Hodges1997,Ojamie1994,Pedulla1996,Cui2006}

Recent work has focused on improving empirical models through inclusion of three-body interactions,
leading to the development of the E3B model.\cite{Kumar2008,Tainter2011} 
Although the inclusion of explicit 3B interactions greatly improves the accuracy of the E3B model relative to
pairwise force fields, the use of rigid water monomers and empirical parameterization necessarily 
misses some of the fundamental properties of the many-body expansion. For example, recent E3B simulations of the 
isomeric equilibria of the water hexamer have led to predictions that are markedly 
different from \ai calculations. Specifically, the prism structure, which corresponds to the energetically 
lowest-lying isomer at the MP2 and CCSD(T) levels of theory, 
\cite{Gora2011,Bates2009a} is unstable in the E3B calculations.\cite{Tainter2012}

Since non-pairwise additive intermolecular interactions
arise primarily from electronic polarization at long distances, several methods
have been proposed to incorporate this effect into the framework 
of classical force fields.\cite{Lopes2009} 
One common approach is the Applequist polarizable point dipole model,\cite{Applequist1972} which
was elaborated upon by Thole to address the so-called polarization
catastrophe.\cite{Thole1981} 
Thole-type polarizable force fields for water include 
TTM3-F\cite{Fanourgakis2008a}, TTM4-F\cite{Burnham2008}, and
AMOEBA\cite{Ren2003} models.

Among methods that attempt to solve directly the many-body problem from
``first principles'', semiempirical models represent 
an attractive alternative due to their computational efficiency. 
Semiempirical models such 
as PM3\cite{Stewart1989} and PM3-MAIS\cite{Bernal-Uruchurtu2000a} were derived 
within the MNDO scheme and differ primarily in the form of the 
core-core repulsion as well as in the precise values of their adjustable parameters.
These models were parametrized either by fitting experimental data for a 
wide variety of systems (PM3) or \ai reference data in the case of PM3-MAIS. 
Due to the use of a minimal basis and the explicit neglect of correlation, semiempirical 
methods are particularly limited in their ability to describe non-bonded interactions. 
This deficiency has been addressed by the SCP-NDDO model, which augments traditional semiempirical methods 
with classical polarization.\cite{Chang2008} SCP-NDDO has shown success in modeling 
water clusters and has recently been extended to simulations of bulk properties.\cite{Murdachaew2011a}

Different DFT methods has also been extensively
used to the study of condensed phases, primarily 
through the use of GGA functionals such as BLYP\cite{Becke1988,Lee1988} and PBE.\cite{Perdew1996} 
However, common density functionals are by construction limited in their ability 
to describe weakly interacting van der Waals complexes. One attempt to address this problem involves 
the addition of a dispersion correction to the energy through the ${C_6}\big/{R^6}$ term, where the 
$C_6$ parameters are atom and basis-set specific \cite{Grimme2004,Grimme2006}. 
These ``DFT-D'' models have successfully described systems such as the 
solvation of iodide in water \cite{Fulton2010}, but are limited by the 
need to develop parameters for each functional/basis set.\cite{Ma2012} 
Furthermore, because the correction is pairwise additive, 
it neglects higher-body dispersion contributions. 
Recent work to address this limitation has been reported.\cite{Grimme2010}

A promising alternative to the pairwise DFT-D correction is represented by the 
non-local van der Waals (nl-vdW) functionals.\cite{Dion2004,Lee2010,Vydrov2010} 
These nl-vdW functionals utilize the electron density 
to define a non-local correlation contribution to the exchange-correlation functional,
leading to a consistent description of both short-range and 
long-range interactions. 
Since no atomic or basis-set dependent parameters are required to describe the dispersion interaction due to 
the explicit dependence of the non-local correlation on the electron density,
nl-vdW functionals, in principle, require minimal parameterization and are system-independent. 
In practice, great care must be taken to avoid double counting of correlation effects
in the combination of semi-local and non-local terms. Van der Waals density functionals 
have recently been applied to the study of liquid water\cite{Wang2011e} and ice\cite{Murray2012}.
 
One final class of models is represented by the \ai-based interaction potentials. 
These models are built upon a rigorous treatment of the many-body expansion of 
interactions and are characterized by having 
a functional form that is sufficiently flexible to accurately map high-quality \ai reference data. 
Examples of such models are DPP2,\cite{Kumar2010} CC-pol,
\cite{Bukowski2008,Bukowski2008a} and WHBB.\cite{Wang2011b}
DPP2 and CC-pol are restricted to the rigid, vibrationally averaged
monomer geometries, while WHBB uses
permutationally invariant polynomials to represent the flexible monomer 
2B and 3B potential energy surfaces (PESs). 
Such \ai-based interaction potentials are quite computationally demanding and 
are most commonly used in calculations
for gas phase systems\cite{Wang2012b}, although bulk properties have been obtained 
from classical simulations with CC-pol\cite{Bukowski2008,Bukowski2007} and DPP2 \cite{Kumar2010}.
Very recently, a flexible version of CC-pol has been 
developed, CC-pol-8s\textit{f}, and the effects of flexibility on the dimer 
vibrational-rotation-tunneling (VRT) spectra have been 
characterized.\cite{Leforestier2012} It was found that both CC-pol-8s\textit{f} and 
HBB2 (the 2B potential of WHBB) reproduce the experimental VRT 
spectra ``about equally well''.\cite{Leforestier2012}

\subsection{\label{sec:2-2}Comparison to CCSD(T)}
\begin{figure*}
\iffigures
\centering
\includegraphics[width=17.0cm]{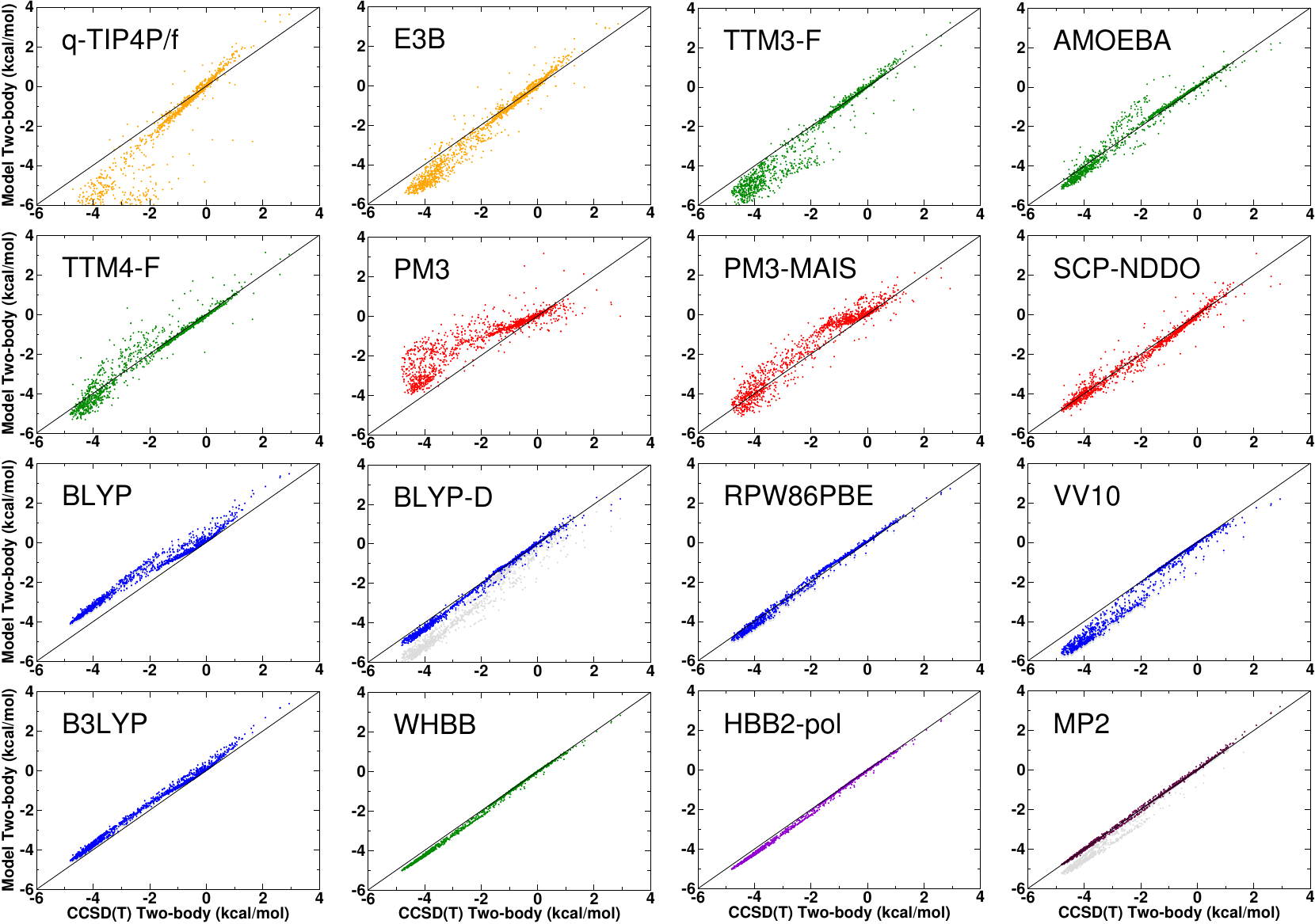}\\
\fi
\caption{\label{fig:e2b}Correlation plots for the 2B interactions. 
Plotted on the x-axes is the BSSE-corrected CCSD(T)/aug-cc-pVTZ energies. On the 
y-axes are the energies for each model. Empirically parametrized models are in orange, 
polarizable models in green, semiempirical methods in red, 
DFT methods in blue, and MP2 in maroon. For DFT and MP2, the 
colored dots are BSSE-corrected energies, while gray dots are BSSE-uncorrected energies. 
The new \ai-based model, HBB2-pol, is in violet.}
\end{figure*}
\begin{figure*}
\iffigures
\centering
\includegraphics[width=17.0cm]{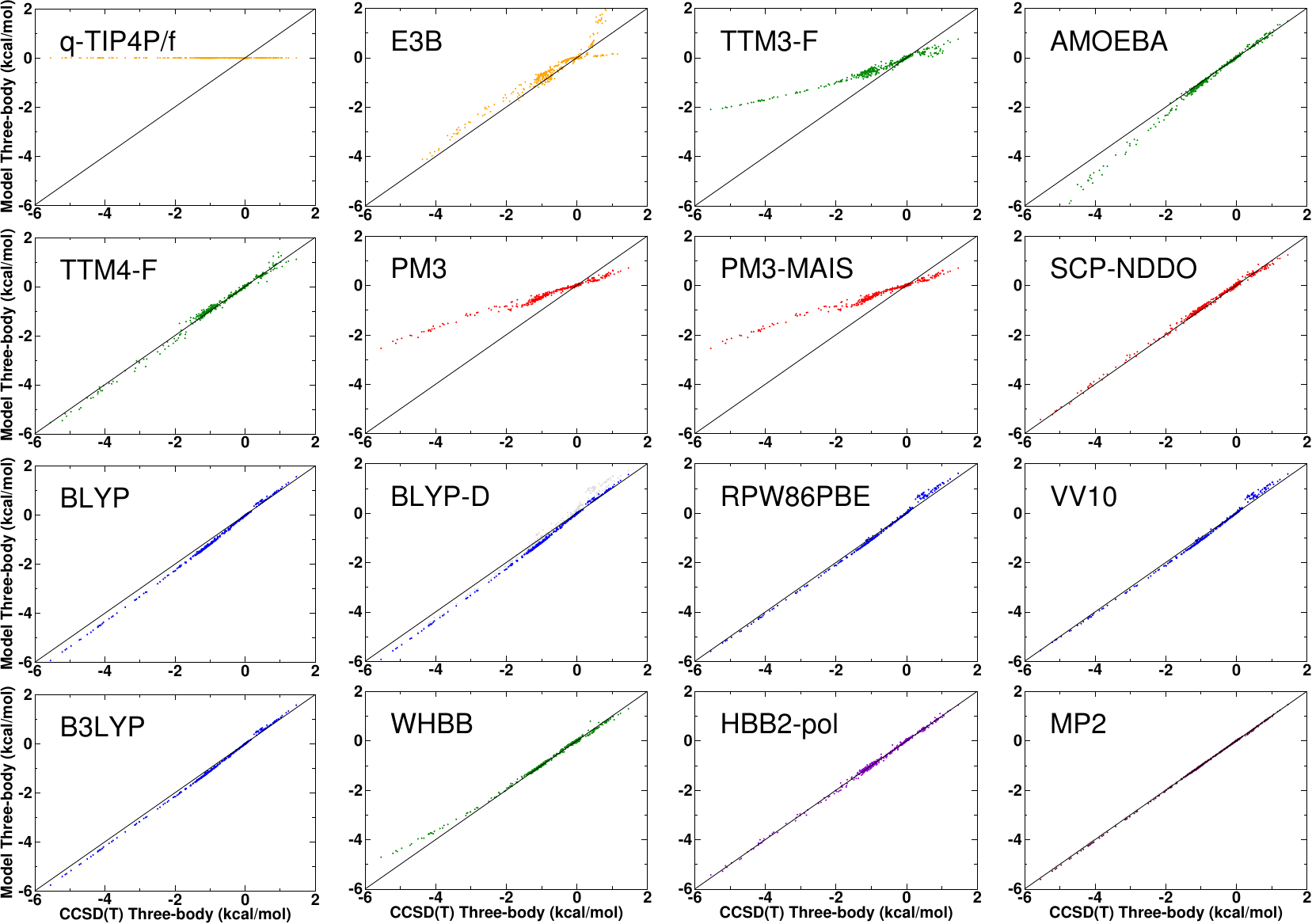}\\
\fi
\caption{\label{fig:e3b}Correlation plots for the 3B interactions. Plotted on the 
x-axes is the CCSD(T)/aug-cc-pVTZ energies corrected for BSSE. On the y-axes are 
the 3B energies for each model. The color scheme is the same as in Figure \ref{fig:e2b}.}
\end{figure*}

Here, we assess the ability of the models presented in 
Section \ref{sec:2-1} to describe the 2B and 3B water interactions.
Roughly 1400 2B interactions and 500 3B interactions were evaluated at the 
CCSD(T)/aug-cc-pVTZ level\cite{Raghavachari1989,Dunning1989} and corrected for 
the basis set superposition error (BSSE) using the counterpoise method.\cite{Boys1970} 
These (flexible) molecular configurations were extracted from 1) classical molecular dynamics (MD) simulations of 
hexamers at T $\leq$ 30~K using the WHBB potential, 2) classical MD simulations of ice I$_\text{h}$ 
carried out with TTM3-F at 50~K, 
and 3) classical MD simulations of bulk water at 298~K and experimental density using TTM3-F. 
Hereafter, these configurations are referred to as ``low-energy'' configurations. 
For the analysis of E3B, the CCSD(T) reference interaction energies were recomputed for ``rigidified'' molecules 
corresponding to the flexible configurations that were used in the comparison of the other models.
All DFT energies were computed using the aug-def2-TZVPP
 basis\cite{Dunning1989,Weigend2005} with the exception of BLYP-D, for which 
 the TZVPP basis was used as in the original parametrization of the model.\cite{Schafer1994,Grimme2004,Grimme2006} 
 MP2 energies were computed with the aug-cc-pVTZ basis, and both DFT and 
 MP2 interactions were corrected for BSSE. All \ai calculations were performed using the 
freely-available \ai package ORCA\cite{Wennmohs2008}. PM3 and PM3-MAIS energies were 
calculated using the AMBER/SQM semi-empirical package\cite{Walker2008}, while the 
SCP-NDDO energies were obtained using CP2K.\cite{cp2k,Murdachaew2011} 
A linear regression analysis for the data 
presented in Figures \ref{fig:e2b} and \ref{fig:e3b}, as well as root mean square error with respect to 
CCSD(T) data, are presented in the supporting material.

Figures \ref{fig:e2b} and \ref{fig:e3b} show correlation plots for the 2B and 3B interactions calculated for all models 
described in Section \ref{sec:2-1} relative to the CCSD(T)/aug-cc-pVTZ energies.
While most empirical pairwise force fields implicitly 
include nuclear quantum effects, models such as q-TIP4P/f and q-SPC/Fw were specifically parameterized
for quantum simulations and, therefore, presumably provide an approximation to the actual Born-Oppenheimer 
PES.\cite{Habershon2009, Paesani2006} As can be seen from Figure 
\ref{fig:e2b}, q-TIP4P/f deviates substantially from the CCSD(T) 2B potential energy surface 
to compensate for the neglect of higher-order interactions (Figure \ref{fig:e3b}). 
Force fields that account for higher-order terms generally provide a
more accurate description of the 2B interactions than effective pairwise models.
In this context, while E3B and TTM3-F/TTM4-F/AMOEBA treat higher-order interactions using different schemes, 
all four models give 2B interactions that are in closer agreement 
with the CCSD(T) results than the effective pairwise models.
It is interesting to note that E3B, which does not not explicitly include induction and was not
parameterized using \ai data, describes the 3B contributions energies 
reasonably well.

The three polarizable models considered in this study (TTM3-F, TTM4-F, and AMOEBA) 
differ in the way they describe the variation of the molecular charge distribution. 
As an isolated monomer deforms, the molecular dipole moment varies in a ``nonlinear'' fashion with respect
to the intramolecular coordinates, resulting in a ``nonlinear dipole moment surface'' (DMS).\cite{Burnham2002c} 
In TTM4-F, the first-order changes of the DMS are fit to electric 
multipoles and polarizabilites calculated at the MP2 level. 
The intramolecular dependence of the atomic charges in TTM3-F was instead
motivated by the observation that, while the gas phase monomer charges decrease during the 
homolytic dissociation, a water molecule in the condensed phase dissociates into charged ions. 
This argument was used to justify an empirical correction to \ai-derived values, giving rise to effective charges that 
increase as the monomer geometry departs from equilibrium. 
By contrast, although AMOEBA takes into account intramolecular flexibility, 
the monomer charges are geometry independent.\cite{Ren2003}
Interestingly, while the accurate monomer DMS has been reported 
to be essential to reproducing the solvated monomer geometry,\cite{Burnham2002c} 
the three-body interaction of AMOEBA is only slightly less accurate than TTM4-F, 
with RMS errors of 0.22 and 0.09 kcal/mol respectively. 
It is also important to mention that, unlike in TTM3-F, in both AMOEBA and TTM4-F 
the molecular polarzability is anisotropic. It is unclear whether the inaccuracy 
observed in the TTM3-F 3B energies arises 
from its use of effective charges, isotropic molecular polarizability or both.

Among the semiempirical methods, PM3 was fitted to a wide range of experimental and \ai data, while 
PM3-MAIS and SCP-NDDO were both fitted to \ai reference data of water clusters. It is therefore not 
surprising that the 2B interactions of PM3-MAIS and SCP-NDDO are in better agreement with
the CCSD(T) data than PM3. It is 
interesting, however, that SCP-NDDO shows much tighter correlation to the \ai data
than PM3-MAIS, even though the latter uses almost twice as many adjustable parameters as SCP-NDDO.
While the two MNDO-type semiempirical methods display
significant deficiencies in describing the 3B interactions, 
SCP-NDDO reproduces the CCSD(T) data quite accurately. 
These results suggest that the addition of classical polarization, as implemented in SCP-NDDO, 
can allow semiempirical methods to accurately describe intermolecular interactions without
requiring extensive reparametrizations of the core-core terms. 

At the 2B level, the GGA density functionals differ appreciably from the CCSD(T) results (see 
Supporting Information for PBE and PBE0 results),
with BLYP systematically underestimating the interaction strength. 
The inclusion of the dispersion correction in BLYP-D improves the agreement with the CCSD(T) values
for the 2B interactions. However, although DFT is less sensitive to basis set incompleteness than wavefunction methods, 
the absence of diffuse functions in the BLYP-D basis results in a large BSSE correction 
(see figure \ref{fig:e2b}, where blue circles give the 
BSSE-corrected interaction and gray circles the BSSE-uncorrected interaction). Indeed, BSSE is so small
for BLYP, B3LYP, RPW86PBE, and VV10 that it is barely visible in Figures~\ref{fig:e2b} and~\ref{fig:e3b}.
While BLYP-D can accurately describe the 2B interactions
when a sufficiently large basis set is used or the energy values are corrected for BSSE, how to balance these 
factors in condensed phase simulations is not straightforward and is the subject of 
ongoing research.\cite{VandeVondele2007,Ma2012}

While the use of hybrid functionals, such as B3LYP \cite{Becke1993,Stephens1994}, results in a much 
tighter correlation to the CCSD(T) data than GGA functionals, B3LYP nonetheless inherently suffers from 
inadequate treatment of dispersion interactions, which leads to an incorrect long-range behavior.\cite{Klimes2012} 
Among recent nl-vdW functionals, VV10 appears to over-correct its parent functional, RPW86PBE, 
leading to over bound 2B interactions. All DFT methods perform reasonably well for the 3B interactions. 
It is important to note that, because the dispersion correction is pair additive, BLYP and BLYP-D 
provide identical 3B interactions. By contrast, nl-vdW functionals include a three-body dispersion correction, 
although this is almost negligible for VV10 (see Supporting Information). MP2 agrees well with CCSD(T), 
with an RMS of  
0.03 and 0.02 kcal/mol for the 2B and 3B interactions, respectively. Consistent with 
previous observations, the magnitude of BSSE is much smaller for 3B than 2B interactions.\cite{Chen1996}

With the exception of MP2, WHBB provides the lowest RMS for the 2B interactions. 
WHBB employs a permutationally invariant polynomial with 5227 coefficients that
were fit to reproduce $\sim$30000 CCSD(T)/aug-cc-pVTZ 2B interactions. 
To account for basis-set truncation, the 
reference 2B interactions were chosen as weighted averages of BSSE-corrected and BSSE-uncorrected 
CCSD(T) interactions.\cite{Shank2009}. 
Since both WHBB and CC-pol reproduce the VRT spectrum of the water dimer with comparable accuracy,\cite{Leforestier2012} 
 a similar agreement with the CCSD(T) data at the 2B level is also expected for CC-pol. 
The agreement of WHBB with the CCSD(T) values for the 3B interactions is
less satisfactory, with WHBB increasingly underestimating the energies of the lowest-lying trimers.
Results for the HBB2-pol model will be discussed in the following sections.

\section{\label{sec:methods} Methods}
Due to its rapid convergence for water, the many-body expansion of interaction energies provides 
a viable way to ``scale up" the CCSD(T) level of accuracy to a large number of molecules. 
Furthermore, by accurately fitting the 1B, 2B, and 3B interactions into a relatively inexpensive function, 
simulations of condensed phases at an effective CCSD(T) level of accuracy become feasible. 
For flexible monomers, the most sophisticated effort along these lines,
WHBB,\cite{Wang2011b} has indisputably proven this concept.
However, WHBB is not directly applicable to bulk phase simulations due to its
prohibitively expensive 3B term. Motivated by this observation, this section reports the development of a new model, HBB2-pol model, beginning with a discussion of the 3B interaction.
\subsection{Three-body Interaction}
Our development exploits the fact that the 3B interaction
in water arises primarily from induction, with all other contributions
vanishing quickly as the intermolecular separation increases.\cite{Mas2003,Chen1996} 
This naturally leads to the following ansatz:
\begin{equation}
  \label{eq:HBB2-pol-3B}
  V_{\mathrm{HBB2-pol}}^{3B} = s_3\polythree + \indthree,
\end{equation}
that represents the 3B interaction as the sum of an induction
term, $\indthree$, and a short-range ``correction", $\polythree$. The physical origins of $\polythree$
are related to the breakdown of the assumptions made in the
derivation the Thole-type induction term as well as to the quantum-mechanical contributions
associated with 3B exchange-repulsion and charge transfer.\cite{Mas2003,Chen1996,Caldwell1990} 
The induction scheme of TTM4-F is used in \indthree due to its 
superior accuracy with respect to other polarizable models (see Fig. \ref{fig:e3b}).
The short-ranged nature of the ``correction" is enforced explicitly by the
switching function, $s_3$,
\begin{equation}
  \label{eq:switch-3B}
  s_3 = f(\xi_{12})f(\xi_{13})
    + f(\xi_{12})f(\xi_{23})
    + f(\xi_{13})f(\xi_{23}),\;\;
\end{equation}
where 
\begin{equation}
  f(\xi) =
    \begin{cases}
      1 & \xi \leq 0 \\
      1 - 3\xi^2 + 2\xi^3 & 0 < \xi \leq 1 \\
      0 & 1 < \xi
    \end{cases}\;,
\end{equation}
\textbf{}\\
$\xi_{ij} = (R_{ij} - R_I)\big/(R_F - R_I)$,  $R_{ij} = \left|\bm{r}_i^{\mathrm{O}} - \bm{r}_j^{\mathrm{O}}\right|$, and $\bm{r}_n^{\mathrm{O}}$ 
denotes the position of the $n$-th molecule
oxygen atom.
This form was found to be better capable of including all trimers in the first solvation shell of a central 
water molecule (in particular, the ``linear'' trimers) than those based on maximum oxygen-oxygen separations. 
Importantly, while the switch in Equation (\ref{eq:switch-3B}) goes 
from 3 to 0 as the trimer passes from the short-range to the long range, the product $s_3 \polythree$ is fitted, 
rather than $\polythree$ by itself. This ensures that no artifact is introduced due to the switching. However, 
because $s_3 \polythree$ is fitted in the context of $\indthree$, this also implies that, unlike in the 
case of the WHBB 3B polynomial, $\polythree$ of HBB2-pol has no meaning by itself but only as
the sum $s_3 \polythree + \indthree$. 

The ability of the short-range polynomial, $\polythree$, to accurately fit reference data depends largely
on its degree, which also determines the associated numerical cost. Consequently, the large 5th and 6th degree 3B 
polynomials in $\polythree$ of WHBB constitute the most computationally taxing part of the model.
The different representation of the 3B interactions (see Eq. (\ref{eq:HBB2-pol-3B})),
along with the improved description of the induction energies in HBB2-pol
allows for an accurate fit of the CCSD(T) reference data with a lower-degree polynomial. 
Specifically, in HBB2-pol the $\polythree$ part is
a sum of second and third degree symmetrized products of exponentials
of the intermolecular separations
\begin{equation}
  \label{eq:variables}
  \eta_{ij} = \exp\big(-k |\bm{r}_i - \bm{r}_j|\big). 
\end{equation}
where $k$ is an adjustable parameter. 
Neither intramolecular distances
 nor the two-body terms
-- those which do not depend on the positions of all three molecules
simultaneously -- were included into the $\polythree$ in HBB2-pol.
Labeling the three molecules as a, b, and c, there
are 27 distances that contribute:
\newcommand{\Hx}[2]{\mathrm{H{#1}}{#2}}
\newcommand{\Ox}[1]{\mathrm{O{#1}}}
\newcommand{\kXX}[1]{k_{\mathrm{#1}}}
\newcommand{\etaX}[3]{\mathrm{e}^{\textstyle -{#1}d\left({#2},{#3}\right)}}
\begin{widetext}
\begin{align*}
  \eta_{1}   = \etaX{\kXX{HH}}{\Hx{a}{1}}{\Hx{b}{1}},\;
  \eta_{2}  = \etaX{\kXX{HH}}{\Hx{a}{1}}{\Hx{b}{2}}&,\;
  \eta_{3}   = \etaX{\kXX{HH}}{\Hx{a}{1}}{\Hx{c}{1}},\\
  \eta_{4}   = \etaX{\kXX{HH}}{\Hx{a}{1}}{\Hx{c}{2}},\
  \eta_{5}  = \etaX{\kXX{HH}}{\Hx{a}{2}}{\Hx{b}{1}}&,\;
  \eta_{6}   = \etaX{\kXX{HH}}{\Hx{a}{2}}{\Hx{b}{2}},\\
  \eta_{7}   = \etaX{\kXX{HH}}{\Hx{a}{2}}{\Hx{c}{1}},\;
  \eta_{8}  = \etaX{\kXX{HH}}{\Hx{a}{2}}{\Hx{c}{2}}&,\;
  \eta_{9}   = \etaX{\kXX{HH}}{\Hx{b}{1}}{\Hx{c}{1}},\\
  \eta_{10}  = \etaX{\kXX{HH}}{\Hx{b}{1}}{\Hx{c}{2}},\;
  \eta_{11} = \etaX{\kXX{HH}}{\Hx{b}{2}}{\Hx{c}{1}}&,\;
  \eta_{12}  = \etaX{\kXX{HH}}{\Hx{b}{2}}{\Hx{c}{2}},\\
  \eta_{13}  = \etaX{\kXX{OH}}{\Ox{a}}{\Hx{b}{1}},\;
  \eta_{14} = \etaX{\kXX{OH}}{\Ox{a}}{\Hx{b}{2}}&,\;
  \eta_{15}  = \etaX{\kXX{OH}}{\Ox{a}}{\Hx{c}{1}},\\
  \eta_{16}  = \etaX{\kXX{OH}}{\Ox{a}}{\Hx{c}{2}},\;
  \eta_{17} = \etaX{\kXX{OH}}{\Ox{b}}{\Hx{a}{1}}&,\;
  \eta_{18}  = \etaX{\kXX{OH}}{\Ox{b}}{\Hx{a}{2}},\\
  \eta_{19}  = \etaX{\kXX{OH}}{\Ox{b}}{\Hx{c}{1}},\;
  \eta_{20} = \etaX{\kXX{OH}}{\Ox{b}}{\Hx{c}{2}}&,\;
  \eta_{21}  = \etaX{\kXX{OH}}{\Ox{c}}{\Hx{a}{1}},\\
  \eta_{22}  = \etaX{\kXX{OH}}{\Ox{c}}{\Hx{a}{2}},\;
  \eta_{23} = \etaX{\kXX{OH}}{\Ox{c}}{\Hx{b}{1}}&,\;
  \eta_{24}  = \etaX{\kXX{OH}}{\Ox{c}}{\Hx{b}{2}},\\
  \eta_{25}  = \etaX{\kXX{OO}}{\Ox{a}}{\Ox{b}},\;
  \eta_{26} = \etaX{\kXX{OO}}{\Ox{a}}{\Ox{c}}&,\;
  \eta_{27}  = \etaX{\kXX{OO}}{\Ox{b}}{\Ox{c}},\;
\end{align*}
\end{widetext}
where $d(\mathrm{X},\mathrm{Y})$ stands for the distance between atoms
X and Y (see also Eq.(\ref{eq:variables})). The monomials
were constructed by symmetrizing the products of the $\eta_n$ variables
with respect to the permutations of both the molecules and the hydrogen
atoms within each molecule (48 elements in the permutation group total).
A total of 131 different monomials were identified (13 are
of second degree, and the remaining 118 are of third degree):
\begin{align*}
  \kappa_{1} &= \eta_{5}\eta_{7} + \eta_{7}\eta_{9} + \eta_{10}\eta_{4}
              + \eta_{5}\eta_{9} + \eta_{6}\eta_{7} + \eta_{11}\eta_{2}\\
              &+ \eta_{2}\eta_{4}
  + \eta_{6}\eta_{8} + \eta_{2}\eta_{3} + \eta_{12}\eta_{4}
    + \eta_{11}\eta_{7} + \eta_{3}\eta_{9} \\
    &+ \eta_{11}\eta_{3} + \eta_{1}\eta_{3}
  + \eta_{5}\eta_{8} + \eta_{12}\eta_{2} + \eta_{10}\eta_{5}
    + \eta_{1}\eta_{4} \\
   &+ \eta_{1}\eta_{9} + \eta_{12}\eta_{8}
    + \eta_{11}\eta_{6} + \eta_{12}\eta_{6} 
    + \eta_{10}\eta_{1} + \eta_{10}\eta_{8},\\
  &\dots\\
  \kappa_{48} &= \eta_{19}\eta_{24}\eta_{6} + \eta_{10}\eta_{13}\eta_{18}
      + \eta_{15}\eta_{22}\eta_{9} + \eta_{10}\eta_{16}\eta_{22} \\
      &+ \eta_{19}\eta_{23}\eta_{7}
     + \eta_{13}\eta_{18}\eta_{7} + \eta_{20}\eta_{23}\eta_{4}
      + \eta_{19}\eta_{23}\eta_{3} \\
      &+ \eta_{11}\eta_{14}\eta_{18}
      + \eta_{19}\eta_{1}\eta_{23}
     + \eta_{12}\eta_{14}\eta_{18} + \eta_{14}\eta_{17}\eta_{4}\\
      &+ \eta_{10}\eta_{16}\eta_{21} + \eta_{16}\eta_{21}\eta_{2}
      + \eta_{19}\eta_{23}\eta_{5}
     + \eta_{11}\eta_{14}\eta_{17} \\
     &+ \eta_{13}\eta_{18}\eta_{8}
      + \eta_{12}\eta_{16}\eta_{22} + \eta_{11}\eta_{15}\eta_{22}
      + \eta_{12}\eta_{14}\eta_{17}\\
     &+ \eta_{15}\eta_{1}\eta_{21} + \eta_{14}\eta_{18}\eta_{7}
      + \eta_{20}\eta_{23}\eta_{8} + \eta_{13}\eta_{17}\eta_{9}\\
     &+ \eta_{15}\eta_{21}\eta_{2}
     + \eta_{13}\eta_{17}\eta_{4} + \eta_{20}\eta_{24}\eta_{4}
      + \eta_{14}\eta_{18}\eta_{8}\\ 
      &+ \eta_{19}\eta_{24}\eta_{7}
      + \eta_{20}\eta_{23}\eta_{5}
     + \eta_{12}\eta_{16}\eta_{21} + \eta_{19}\eta_{24}\eta_{2}\\
      &+ \eta_{13}\eta_{17}\eta_{3} + \eta_{20}\eta_{24}\eta_{2}
      + \eta_{16}\eta_{22}\eta_{5}
     + \eta_{1}\eta_{20}\eta_{23} \\
     &+ \eta_{15}\eta_{22}\eta_{6}
      + \eta_{14}\eta_{17}\eta_{3} + \eta_{20}\eta_{24}\eta_{6}
      + \eta_{15}\eta_{21}\eta_{9}\\
     &+ \eta_{10}\eta_{13}\eta_{17} + \eta_{13}\eta_{18}\eta_{9}
      + \eta_{15}\eta_{22}\eta_{5} + \eta_{11}\eta_{15}\eta_{21}\\
      &+ \eta_{16}\eta_{22}\eta_{6}
     + \eta_{16}\eta_{1}\eta_{21} + \eta_{19}\eta_{24}\eta_{3}
      + \eta_{20}\eta_{24}\eta_{8} \\
  &\dots\\
  \kappa_{131} & = \eta_{25}\eta_{26}\eta_{27}.
\end{align*}
The $\polythree$ itself was then taken as a linear combination of
$\kappa_n$:
\begin{equation}
  \label{eq:V_poly}
  \polythree = \sum\limits_{n=1}^{131}v_n\kappa_n,
\end{equation}
with the coefficients $v_n$ obtained using the least squares fit\cite{Galassi2009}
to the CCSD(T) data. The gradient of $\polythree$ with respect to the
atomic positions was computed using MAPLE.\cite{maple}

\subsection{\label{sec:methods-2}Composition of three-body training set}

After translational and rotational invariance, the 3B interaction
potential for flexible water molecules is 21 dimensional. Since this
high-dimensional PES cannot be readily 
trained on a grid, a training set representative of the ``important regions'' of 
the 3B PES was generated by including: 1) repulsive configurations with positive binding energies
that are compressed
relative to the trimer global minimum, 2) ``low-energy'' configurations
 with thermally accessible binding
energies, and 3) long-range trimer configurations that have weak
3B interactions.
The total training set consists of 8019 trimer configurations for
which the 3B energies were computed at the
CCSD(T)/aug-cc-pVTZ level and corrected for BSSE.\cite{Boys1970}

The majority of the configurations in the training set correspond
to an expanded set of ``low-energy'' configurations, similar in composition
to that used in the analysis of section \ref{sec:2}. 
Of the 5515 thermally accessible configurations that were used in the fit, 996 trimers
were selected from MD simulations of clusters (trimers and hexamers)
at 30~K on the WHBB potential energy surface, 3311 trimers were extracted
from classical MD simulations of hexagonal ice and liquid water carried out with TTM3-F 
at 50~K and 300~K respectively, 792 trimers
were obtained by randomly orienting water monomers in geometries
near the global minimum, and 416 trimers were obtained from scans of low-energy structures. 

The long-range portion of the training set consists of 432 weakly
interacting trimers, with an average O-O distance of at least 5.5~{\AA} between monomers.
After verifying that the CCSD(T)/aug-cc-pVTZ
3B interaction energy for these long-range configurations was in 
agreement with the 3B induction energy from TTM4-F (within
the error associated with basis-set truncation), we assigned these
configurations the TTM4-F 3B induction energy. This enforces 
the ``boundary condition" that the 3B interaction
of HBB2-pol become pure induction at long distances.
\begin{figure}
\iffigures
  \includegraphics[width=7.5cm]{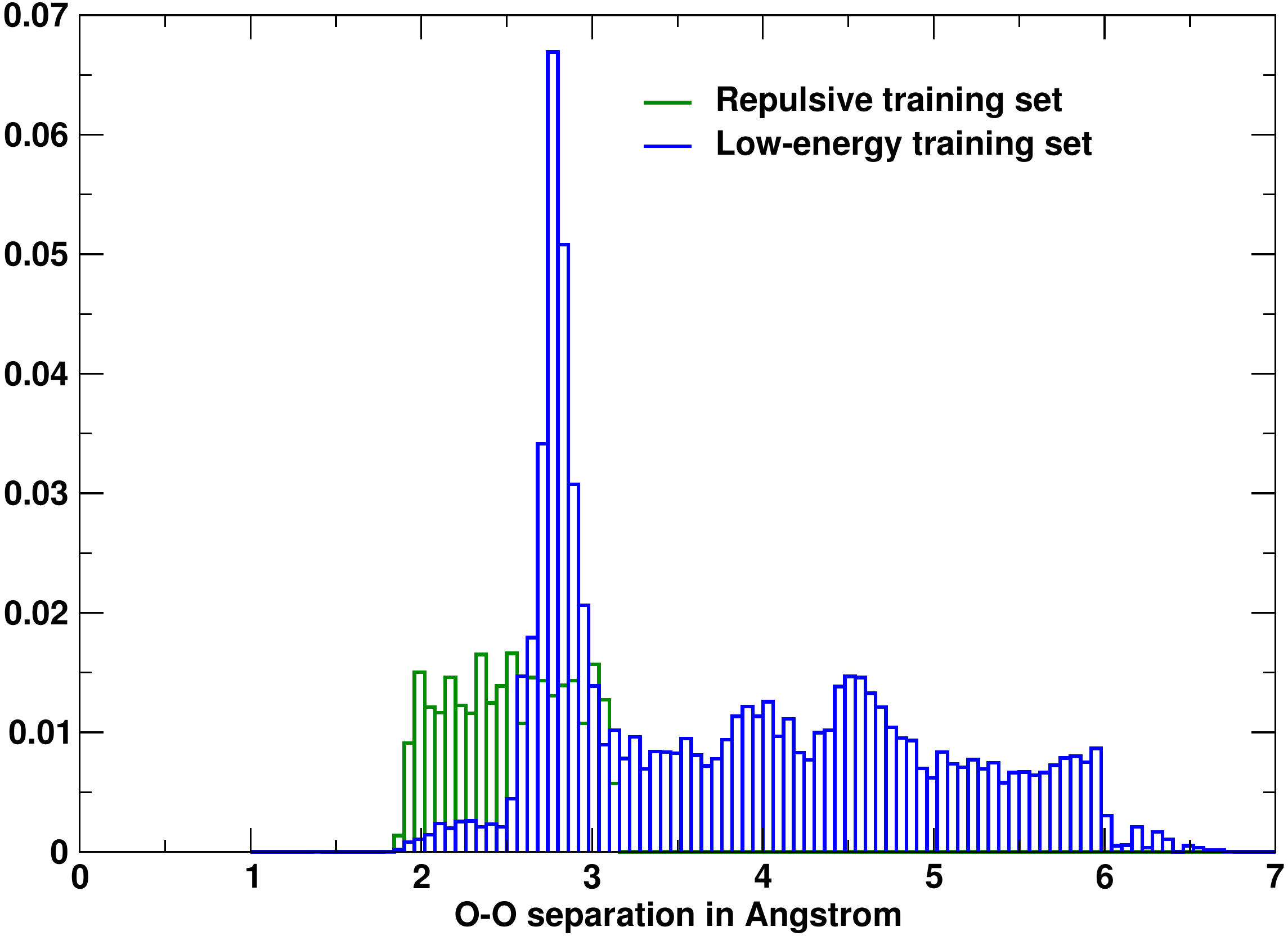}
\fi
  \caption{\label{fig:ts_vs_rOO}Distribution of O-O distances in the three-body training set, 
  including all three O-O distances per trimer. 
  ``Low-energy'' denotes configurations which were thermally accessible (see text for details).}
\end{figure}

It is important to mention that our
initial model were fitted to a training set that emphasized only the low-energy and
long-range regions, which is consistent with the parameterization strategy
adopted for the WHBB 3B potential\cite{Wang2011b}. While the resulting models
succeeded in predicting the relative stabilities of trimer and hexamer 
isomers, we found that they were numerically unstable in MD simulations
of larger clusters, such as the 32-mer. This was related to insufficient
coverage of the repulsive, short-ranged region of the trimer PES. 
\begin{figure}
\iffigures
  \includegraphics[width=7.5cm]{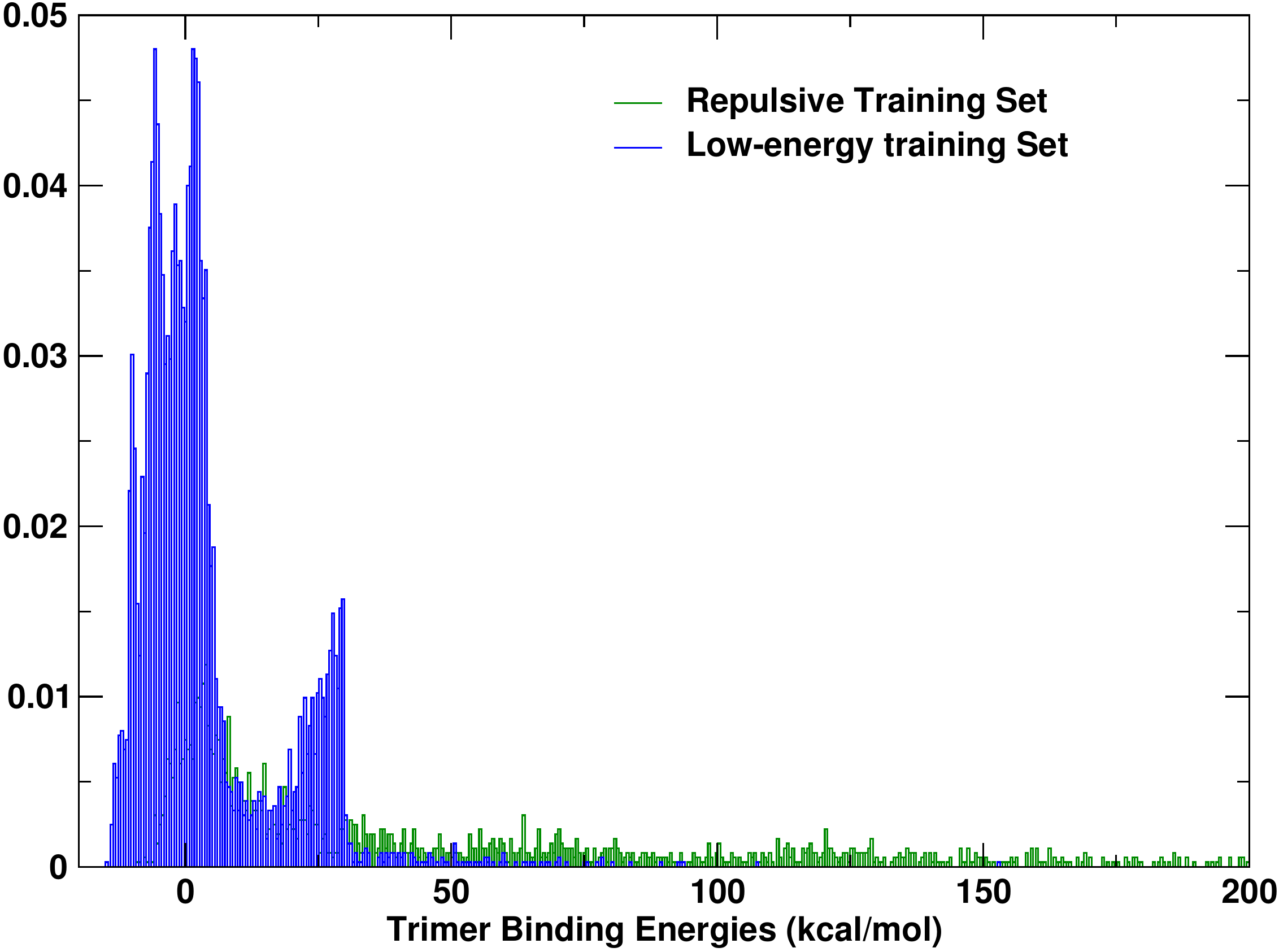}
\fi
  \caption{\label{fig:ts_vs_ebind}Distribution of trimer binding energies in the three-body training 
  set.}
\end{figure}
To address this instability, 2072 configurations were generated by
performing random rotations on monomers that were compressed relative to
the trimer minimum geometry. As shown in 
Figure \ref{fig:ts_vs_rOO}, this repulsive training set evenly covers O-O
distances from 1.8~{\AA} to 3.2~{\AA}. Many of these compressed configurations,
however, correspond to trimers that would practically never be sampled in
simulations of water under ambient conditions. While including 
these configurations in the training set was required to ensure the
numerical stability of the model, it was also necessary to guarantee that these
highly-repulsive configurations did not degrade the quality 
of the fit in the region near the minimum. This consideration
is discussed in the following section.

\subsection{\label{sec:methods-3}Testing the accuracy of the three-body fit}

In order to assess the accuracy of any model with respect to \ai data, it
would be ideal to parameterize the model based on one set of reference
data (the ``training'' set) and validate the model using
a separate set of data (the ``testing'' set). However, due to the large
computational cost associated with BSSE-corrected CCSD(T)/aug-cc-pVTZ 3B
interactions and the fact that the model should be parametrized to as
large of a reference set as possible, having two distinct sets is
unfeasible. In order to estimate the bias associated with training
and testing on the same reference data, the complete training set was equally divided
into two parts, a training set and a testing set. Care was taken
to preserve the relative compositions of each set, e.g., half of the
972 configurations extracted from simulations of ice were randomly
selected and placed in the training set while the other half 
was placed in the testing set. 

Since the goal of HBB2-pol is to describe water from
small molecular clusters in the gas phase to condensed phases, 
we found it important to decompose the RMS into
categories: 1) trimers with a binding energy of less than 5~kcal/mol, which
are particularly important for low-temperature cluster isomer
relative stabilities, 2) trimers with a binding energy of less
than 30~kcal/mol, which are sampled during path-integral molecular
dynamics (PIMD) simulations of bulk water at ambient conditions, and 3) the
complete training set, regardless of binding
energy (see Table \ref{tab:testtrain}).
Due to the proximity
of the energies for different isomers of the small clusters, it is
desirable that 3B interactions corresponding to clusters with the lowest
binding energies have the lowest RMS.
As was discussed in Section \ref{sec:methods-2}, it was necessary to include
3B energies corresponding to trimers with positive binding energies
to ensure the stability of HBB2-pol. At the same time,
care was taken to ensure that the 
accuracy in the low-energy region was not diluted by being needlessly accurate for 
trimers with enormous binding energies that will rarely be sampled. By
appropriately weighting the reference data, HBB2-pol achieves the
best accuracy in the low-energy region, slightly larger RMS in
the intermediate region ($E_{bind} < 5$ kcal/mol), and the largest
error for those configurations with binding energies greater than 30 kcal/mol.
To obtain this RMS distribution, configurations with 3B interactions were 
weighted according to their trimer binding energies, where configurations with $E_{bind} < 15$ 
kcal/mol were given a weight of 1.0, while weights for configurations binding energies 
larger than 15 kcal/mol a weight of 
$e^{-a(E_{bind} - E_{0})}$, where a = 0.05 kcal/mol$^{-1}$ and $E_0$ was 15 kcal/mol.
\begin{table}
\caption{\label{tab:testtrain}RMS deviation of three-body interactions from trimers 
with binding energies less than 5kcal/mol, less than 30kcal/mol, and for the 
complete training set. See main text for details.}
\begin{tabular}{rC{1.5cm}C{1.5cm}C{1.5cm}}
\firsthline\noalign{\smallskip}
\multicolumn{1}{r}{}
 &  \multicolumn{1}{C{1.5cm}}{WHBB}
   &  \multicolumn{1}{C{1.5cm}}{$\indthree$}
      &  \multicolumn{1}{C{1.5cm}}{HBB2-pol} \\
\noalign{\smallskip}\cline{2-4}\noalign{\smallskip}
\multicolumn{1}{r}{}&\multicolumn{3}{c}{RMS for Training Set}\\	
\noalign{\smallskip}\cline{1-4}\noalign{\smallskip}
E$_{bind}$ $<$ 5    & 0.10 & 0.13 & 0.06 \\
E$_{bind}$ $<$ 30   & 0.14 & 0.35 & 0.10 \\
Total               & 0.69 & 2.26 & 0.85 \\
\noalign{\smallskip}\cline{1-4}\noalign{\smallskip}
\multicolumn{1}{r}{}&\multicolumn{3}{c}{RMS for Testing Set}\\	
\noalign{\smallskip}\cline{1-4}\noalign{\smallskip}
E$_{bind}$ $<$ 5    & 0.10 & 0.13 & 0.06 \\
E$_{bind}$ $<$ 30   & 0.15 & 0.35 & 0.16 \\
Total               & 0.68 & 2.20 & 0.82 \\
\noalign{\smallskip}\cline{1-4}\noalign{\smallskip}
\multicolumn{1}{r}{}&\multicolumn{3}{c}{RMS for Complete Set}\\	
\noalign{\smallskip}\cline{1-4}\noalign{\smallskip}
E$_{bind}$ $<$ 5    & 0.10 & 0.13 & 0.05 \\
E$_{bind}$ $<$ 30   & 0.14 & 0.34 & 0.11 \\
Total               & 0.66 & 2.17 & 0.83 \\
\noalign{\smallskip}\lasthline%\noalign{\smallskip}
\end{tabular}
\end{table}

The results presented in Table \ref{tab:testtrain} confirms the ability of the HBB2-pol 3B
function to recover the \ai data and demonstrates that the training
set is sufficiently large to render the model insensitive to the size of the 
training set. This analysis does not, however, probe whether the training
set includes all the physically relevant configurations. Assessing
whether the composition of the training set is biased can only be
accomplished by examining \ai properties such as relative cluster
isomer stabilities, experimental properties such as virial coefficients, and
the overall numerical stability of the model.

\subsection{\label{sec:methods-4}HBB2-pol}
Using the 3B interaction proposed in Eq.~(\ref{eq:HBB2-pol-3B}), HBB2-pol 
has been developed through the many-body expansion, Eq.~(\ref{eq:many-body}).
The spectroscopically-accurate monomer potential
energy surface of Partridge and Schwenke is used for the 1B
terms.\cite{Partridge1997a} For the 2B interaction, the
HBB2 PES\cite{Shank2009} is employed at short-range. 
The HBB2 short-range 2B interaction is smoothly switched to electrostatics/induction
plus dispersion term at long-range over the interval 
$R^{2B}_I=5.5\text{\AA}<R_{\mathrm{OO}}<R^{2B}_F=7.5\text{\AA}$:
\begin{align*}
  V^{2B} &= (1-s_2) \hbbtwo\\ 
         & + s_2\Big[\electwo
                   + \indtwo
           - \frac{C_6}{R_{\mathrm{OO}}^6}\Big],\\
  s_2 &= \begin{cases}
          0 & \xi\leq 0\\
          \xi^3(10 - 15\xi + 6\xi^2) & 0 < \xi \leq 1 \\
          1 & 1 < \xi
        \end{cases},
\end{align*}
where $\xi=(R_{\mathrm{OO}}-R_I)\big/(R_F - R_I)$. The $\electwo$ and $\indtwo$
terms have the same form as TTM4-F, and
the value of $C_6$ was taken as the difference
between the sum of the {\ai} van der Waals constants describing
the dispersion and induction interactions in the asymptotic region
from Ref.~\citenum{Bukowski2008},
and the $1/R_{\mathrm{OO}}^6$ coefficient of the isotropic part of
$\indtwo$
\begin{displaymath}
  C_6 = (47.053232\;\mathrm{a.u.} + 10.66517\;\mathrm{a.u}) - 2\alpha \mu^2,
\end{displaymath}
(using the isotropic molecular polarizability given by TTM4-F,
$\alpha = 1.41567$~{\AA}$^3$ the molecular
dipole $\mu = 1.864047$~D).
The 3B interactions are those presented in Eq.(\ref{eq:HBB2-pol-3B}), and
all the higher-body terms are approximated by the induction energy as in TTM4-F,
\begin{align}
 V^{NB} &= \indn
      - \sum_{i<j<k}^N\indthree(i,j,k)\nonumber\\
              &- \sum_{i<j}^N \indtwo(i,j).
\end{align}
Since \hbbtwo accounts for polarization at the 2B level, the short-range 
2B contribution must be subtracted from the N-body induction to prevent double counting. 
This problem does not arise for the 3B interactions since induction is not modified at the 
3B level (Eq. (\ref{eq:HBB2-pol-3B})).
The HBB2-pol interaction energy for $N$ water molecules is thus given by the
following expression
\begin{align}
     E&_{N-mer} = \sum_i^N \ps(i) \nonumber\\
     &+ \sum_{i<j}^{N}\biggl\{(1-s_{2})
        \bigl[\hbbtwo - \indtwo\bigr]
     + s_{2}\bigl[\electwo
       - \frac{C_6}{R_{\mathrm{OO}}^6}\bigr]\biggr\} \nonumber\\ 
     &+ \sum_{i<j<k}^N s_3 \polythree(i,j,k)
          + \indn(1,\dots,N),
\end{align}
where the 3B switching functions, $s_3$, is given by Eq.(\ref{eq:switch-3B}). 

\section{\label{sec:results}Results}
In this section we demonstrate the ability
HBB2-pol to reproduce CCSD(T) calculations and the experimental 
second and third virial coefficients. 

\subsection{\label{sec:results-scans}
Short-range Three-body Interaction Addresses Systematic
Flaws in Polarizable Models}
\begin{figure}
\centering
\iffigures
    \includegraphics[width=7.5cm]{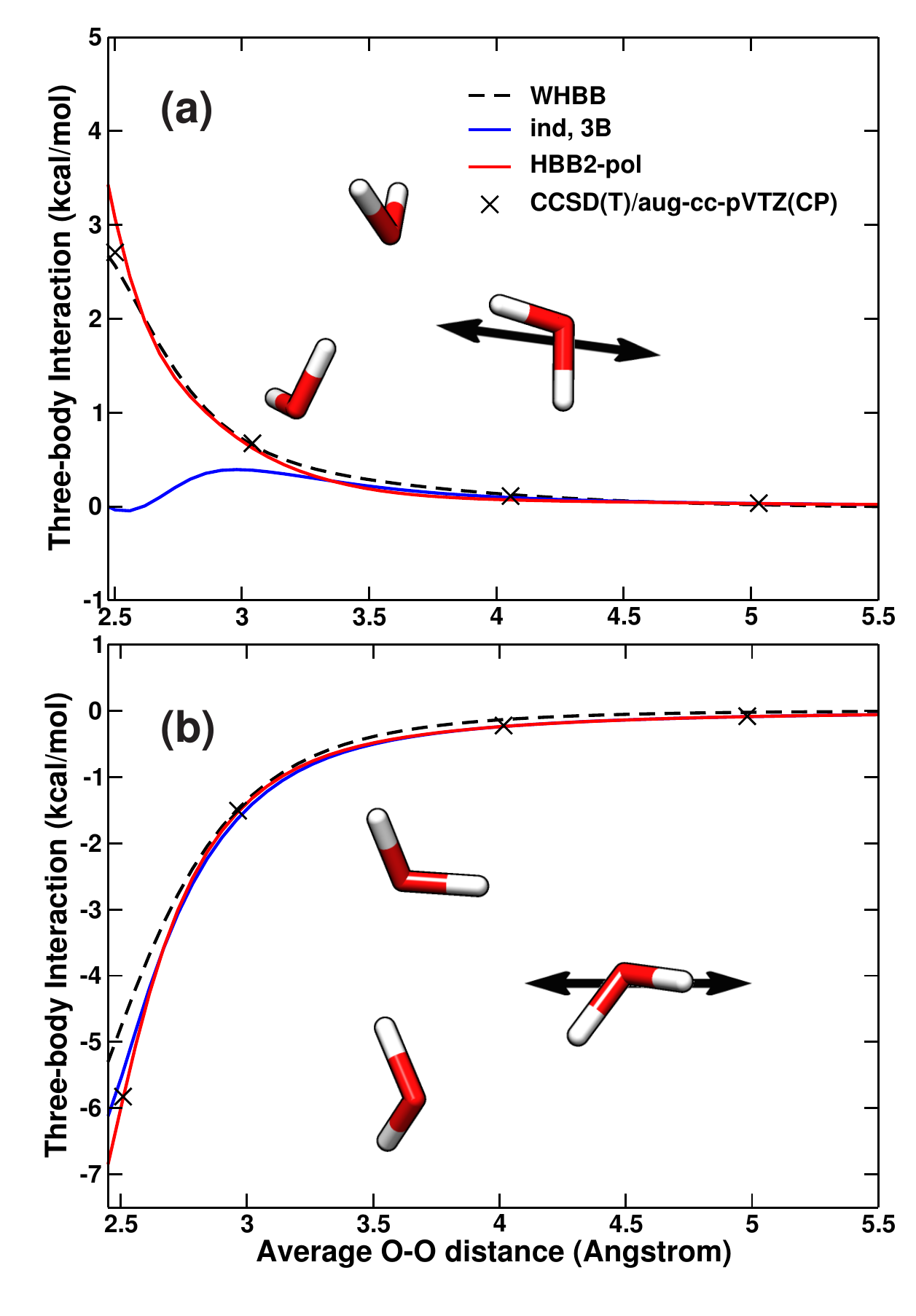}     
\fi
\caption{\label{fig:trimer_scans}
Three-body interaction energy for two cuts
through the water trimer potential energy surface:
WHBB (black dashes), three-body induction (blue), HBB2-pol (red) and
CCSD(T)/aug-cc-pVTZ (crosses).
}
\end{figure}

As discussed above, the three-body interactions primarily originate from
induction, though for more strongly 
bound clusters effects including
exhange-repulsion and charge transfer can also make a significant
contribution.\cite{Mas2003,Chen1996}
As a consequence, force fields which treat only induction
 are inherently unable to fully describe
3B interactions. By contrast, models that only treat short-range 3B interactions 
are unable to describe the induction interactions that dominate at
long range. This is illustrated in Figure \ref{fig:trimer_scans},
where HBB2-pol is compared with WHBB, TTM4-F 3B induction, and CCSD(T)/aug-cc-pVTZ
data along two representative cuts through the water trimer PES. 
The CCSD(T) reference data used for this comparison were not included in the training set. 
This comparison clearly shows that the addition of the short-range ``correction'' 
to the induction brings the 3B interactions of HBB2-pol into close agreement with the CCSD(T) data.

\subsection{\label{sec:results-isomers}
Trimer stationary points
}
To assess the combined accuracy of the 1B, 2B, and
3B interactions of HBB2-pol, we studied the relative
energies of four water trimer isomers identified in
Table~\ref{tab:trimer_isomers} by their free-hydrogen
orientation: ``u'' for pointing up, ``p'' if the hydrogen
lies in the plane of the oxygen atoms, and ``d'' for pointing
down. The energetics of these structures have
been reported in Ref.~\citenum{Anderson2004}, where
geometries optimized at the MP2/aug-cc-pVQZ level were used to calculate
the energies at the CCSD(T) level in the complete
basis limit. The HBB2-pol energies relative to the trimer global
minimum (uud) are reported in Table~\ref{tab:trimer_isomers}
for the geometries optimized on the HBB2-pol PES.
These favorably interacting trimers have CCSD(T) binding energies of
approximately -15 kcal/mol, which implies that the energies separating these stationary
points are on the order of 1-10\% of the binding energy. 
The HBB2-pol relative energies fall within 0.05 kcal/mol of the
reference data for the upd and uuu structures, while a larger difference
of 0.23 kcal/mol is obtained for
the ppp structure. HBB2-pol, however, is not expected to achieve
perfect agreement with the CCSD(T)/CBS data\cite{Anderson2004} due
to the different basis set used in the fit of the 3B terms.

\begin{table}
\caption{\label{tab:trimer_isomers}
Relative energies of water trimer isomers
with respect to the global minimum ``uud" in kcal/mol.
CCSD(T) energies were extrapolated to complete basis set limit, from Ref. \citenum{Anderson2004}.
}
\begin{tabular}{C{2.7cm}C{2.2cm}C{2.2cm}}
\cline{1-3}\noalign{\smallskip}
 \multicolumn{1}{C{2.7cm}}{}
 &  \multicolumn{1}{C{2.2cm}}{CCSD(T)}
 &  \multicolumn{1}{C{2.2cm}}{HBB2-pol}\\
\noalign{\smallskip}\cline{1-3}\noalign{\smallskip}
\iffigures
\includegraphics[width=2.7cm]{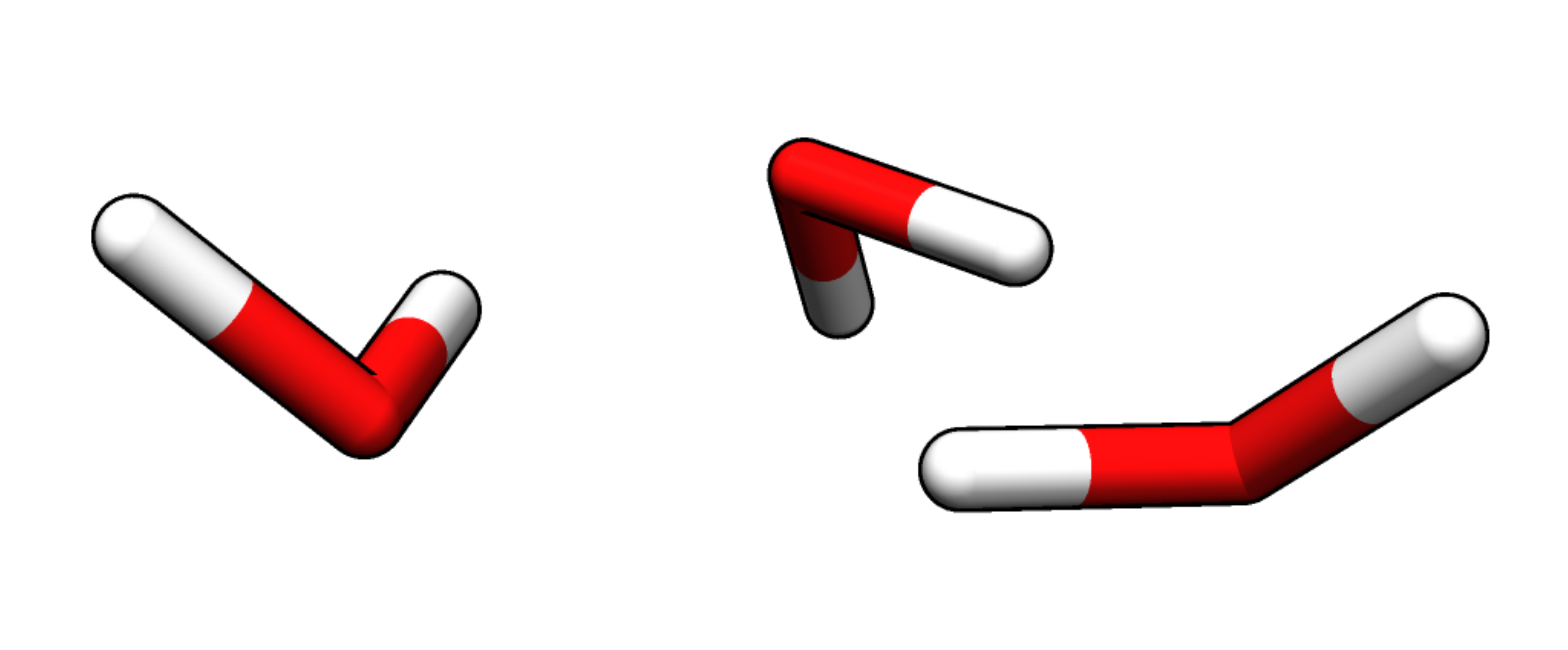}\fi \hspace{0.5cm}  uud & 0.0 & 0.0 \\
\noalign{\smallskip}\cline{1-3}\noalign{\smallskip}
\iffigures\includegraphics[width=2.7cm]{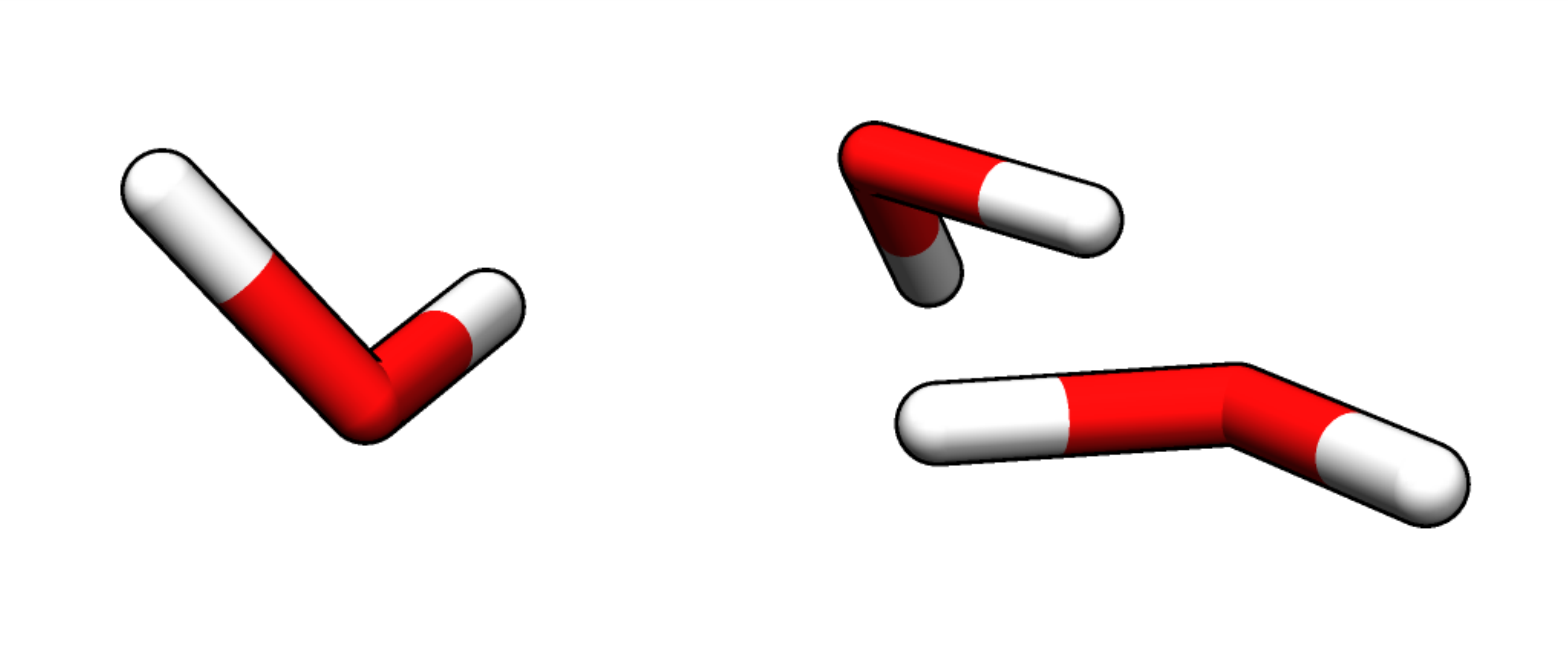}\fi\hspace{0.5cm} upd & 0.23 & 0.18 \\
\noalign{\smallskip}\cline{1-3}\noalign{\smallskip}
\iffigures\includegraphics[width=2.7cm]{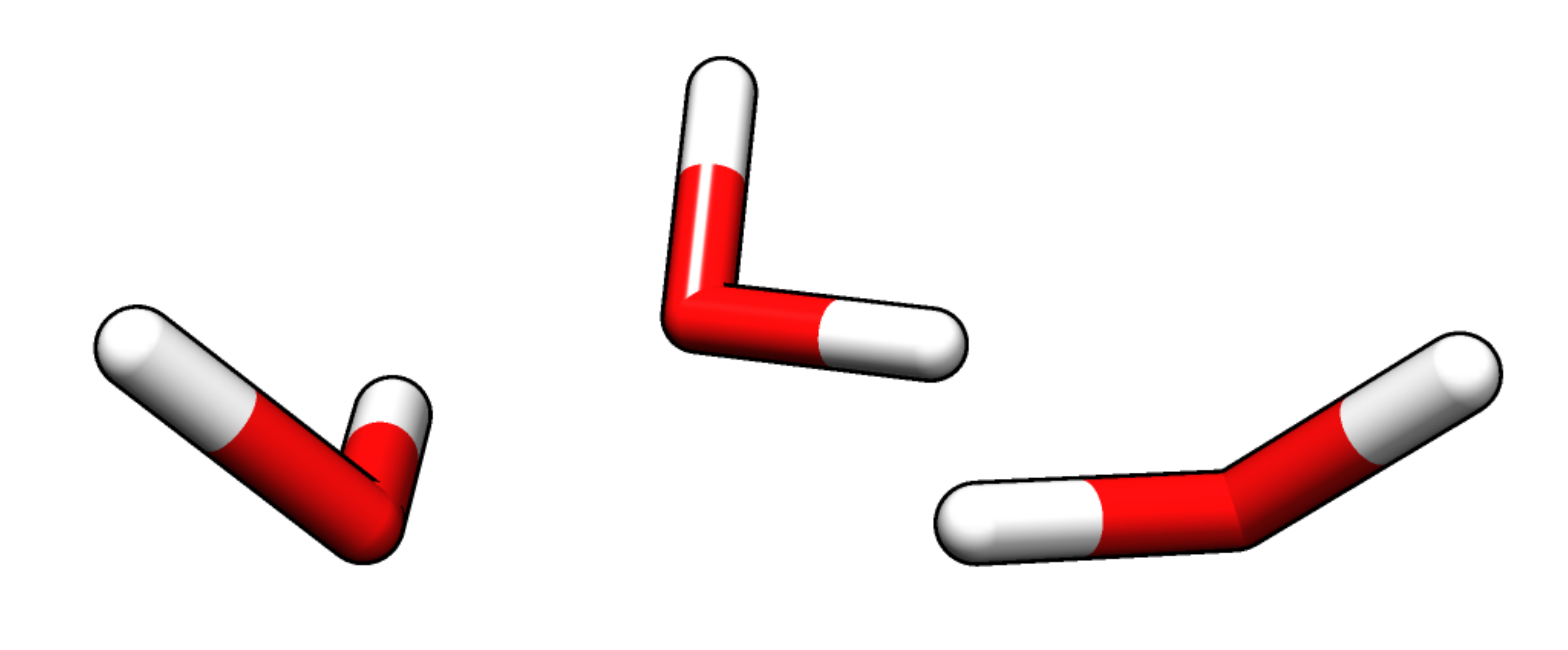}\fi \hspace{0.5cm}  uuu & 0.77 & 0.73\\
\noalign{\smallskip}\cline{1-3}\noalign{\smallskip}
\iffigures\includegraphics[width=2.7cm]{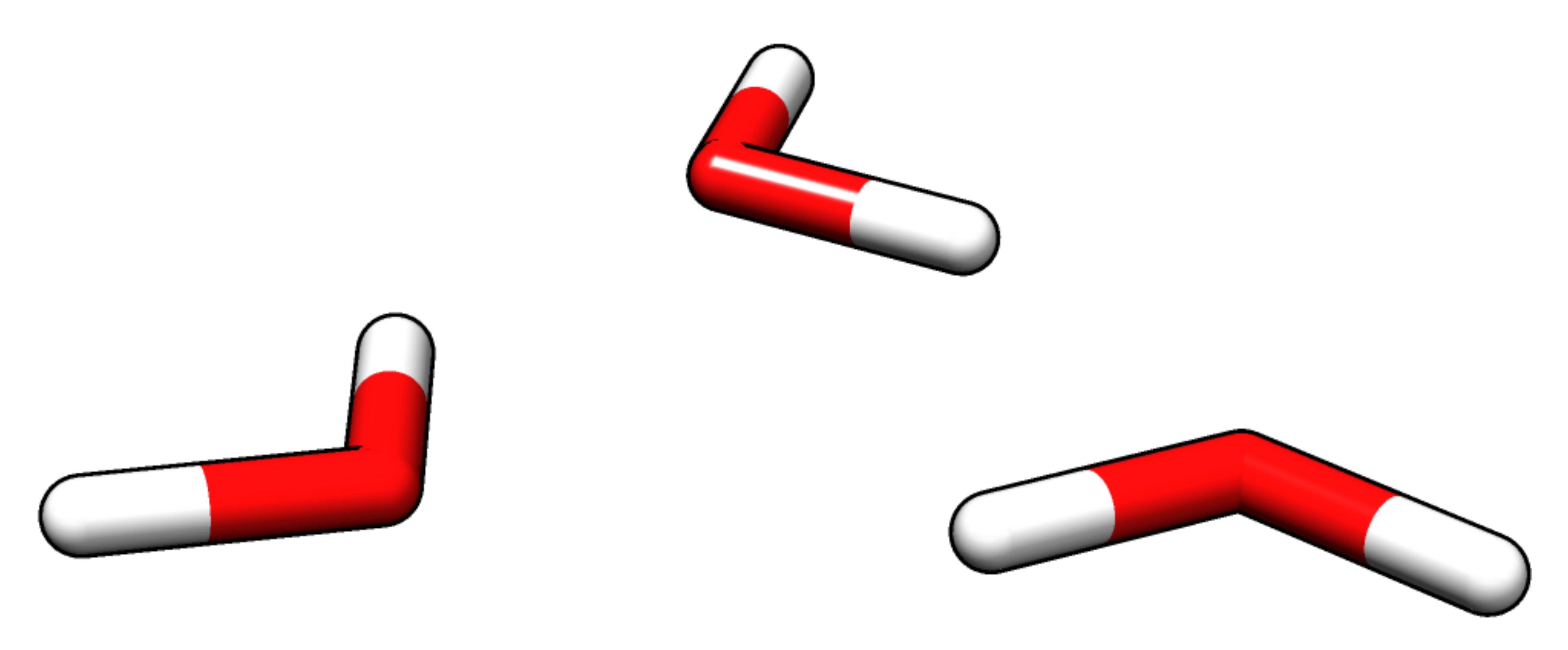}\fi \hspace{0.5cm}  ppp & 1.25 & 1.02 \\
\noalign{\smallskip}\cline{1-3}\noalign{\smallskip}
\end{tabular}
\end{table}

\subsection{\label{sec:results-virial}
Virial coefficients
}
Virial coefficients are derived from the virial equation of state
that expresses $p/k_BT$ as a power series in density and
gauge deviations from the ideal gas behavior,
\begin{equation}
  \label{eq:veos}
  \frac{p}{k_B T} = \frac{N}{V}\biggl[ 1 + B_2 \frac{N}{V}
    + B_3 \biggl(\frac{N}{V}\biggr)^2 + ... \biggr].
\end{equation}
Here, $B_2$ and $B_3$ are the second and third virial
coefficients, respectively.\cite{Hill1986,Mayer1940,Mason1969} 
The second virial coefficient depends only on the pair interaction while 
the third virial coefficient also includes the 3B interaction, but no $(n>3)$-body energies. 
Since both $B_2$ and $B_3$ are
experimentally accessible, the virial coefficients provide a critical assessment
of the accuracy of water potentials. 

Neglecting the contribution of intramolecular vibrational modes
(that is, assuming rigid monomers) and nuclear quantum effects, the second
virial coefficient is given by\cite{Mason1969},
\begin{align}
  \label{eq:virial2}
  B_2(T) &= -2\pi\int\mathrm{d}R_{12}\,R_{12}^2
    \bigl<f_{12}\bigr>_{\bm{\Omega}_1, \bm{\Omega}_2},\nonumber\\
  f_{12}&=\mathrm{e}^{-\beta V^{2B}(R_{12},\bm{\Omega_1},\bm{\Omega_2)}}
       - 1,
\end{align}
where $\beta=1/k_BT$ is the inverse temperature, $R_{12}$ is the
distance between the monomer centers of mass,
$V^{2B}(R_{12},\bm{\Omega_1},\bm{\Omega_2})$
is the intermolecular interaction energy, and the angular brackets
stand for the average over the orientations of the
molecules $\bm{\Omega}_{1,2}$. The Mayer function, $f_{12}$, has
the useful property of going to zero as the molecules move apart.
To numerically evaluate Eq. (\ref{eq:virial2}), the Simpson rule
is used to calculate the radial component of the integral, while
Monte Carlo integration is used to evaluate
the orientational average using $10^5$ random orientations
of the monomers at each point on the radial grid.
\begin{figure*}
        \caption{\label{fig:virial2}Effect of the rigid monomer configuration on the error in the classical 
        second virial coefficient relative to experiment.\cite{Harvey2004} Differences between WHBB 
        and HBB2-pol were indistinguishable on the scale of this plot, so both have been assigned to 
        the red line. For E3B, its rigid monomer geometry was used in both plots.}
\iffigures
\centering
                \includegraphics[width=16cm]{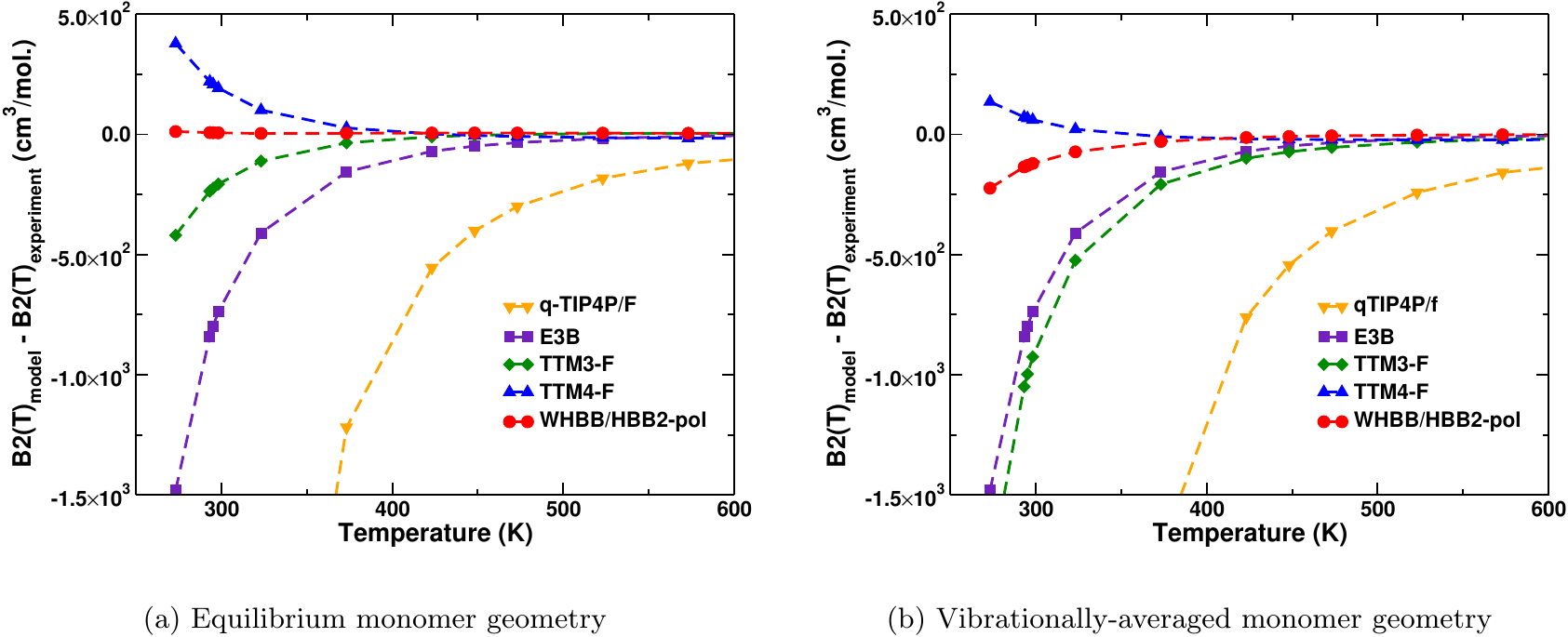}\\
\fi
\end{figure*}

To explore the sensitivity of the second virial coefficient to the
choice of the rigid monomer geometry, its values are calculated
using two different configurations: 
the Born-Oppenheimer minimum energy configuration given by the
Partridge-Schwenke potential energy surface
($r_{OH}^{eq}= 0.95784\text{\AA}$ and $\theta_{HOH}^{eq}= 104.508^{\circ}$) 
\cite{Partridge1997a}, and the ground-state vibrationally-averaged
configuration of $r_{OH}^{eq}=0.9716256 \text{\AA}$ and
reported in
reference \citenum{Mas1996} ($\theta_{HOH}^{eq}=104.69^{\circ}$). 
Importantly, examining the results for these
two configurations provides not only an estimate of the effect of flexibility, 
but also of nuclear quantum effects ``sensed" through the ground-state
vibrationally averaged configuration.
 
Plotted in Figure \ref{fig:virial2} is the difference between the
calculated virial coefficients and the experimental data.\cite{Harvey2004} Since $B_2$ 
is the integral of $e^{-\beta V^{2B}}-1$, comparison
to low-temperature results are 
particularly interesting as these are most sensitive
to the region near the dimer 
minimum geometry. For the Born-Oppenheimer equilibrium geometries (Fig. 
\ref{fig:virial2}a), the \ai-based potential energy
surfaces WHBB and HBB2-pol very closely 
reproduce the experimental data. Since HBB2-pol and WHBB share the same two-body PES at dimer
separations of less than 5.5{\AA}, it is not surprising that both models
 predict similar values for the second virial coefficient. Though detectable, the
differences between WHBB and HBB2-pol are not visible on the scale
of Figure \ref{fig:virial2}, so both have 
been assigned to the red line. 

While both qTIP4P/f and E3B were empirically parametrized, the
inclusion of explicit 3B interactions in E3B allows for a more
accurate 2B interaction. Pairwise-additive models, such as 
qTIP4P/f, on the other hand, rely on the 2B interaction
to (partially) recover 3B effects and would therefore be expected
to have a much larger error for the second virial coefficient. When the 
monomer geometry is changed
from the equilibrium to the vibrationally-averaged geometry, there is a 
small decrease in the value of $B_2(T)$ for most models. 
TTM3-F, however, exhibits a large change in its second virial
coefficient. This is likely due to the empirical modification of
the dipole moment surface, which only affects TTM3-F when
the monomer distorts from the
equilibrium configuration.\cite{Fanourgakis2008a} 

The third virial coefficient depends on the interaction of trimers and
provides an indirect measure of the 3B interaction.
Following Hill\cite{Hill1986}, the third virial 
coefficient can be separated into a pairwise component, $B_3^0(T)$, and
the 3B contribution, $\Delta B_3^{3B}(T)$:
\begin{equation}
  B_3(T) = B_3^0(T) + \Delta B_3^{3B}(T),
\end{equation}
where the pairwise contribution is the integral over the
product of the three Mayer functions in Eq. (\ref{eq:3rdvirial2b}),
and the 3B contribution is given by Eq. (\ref{eq:3rdvirial3b}).
\begin{widetext}
\begin{equation}\label{eq:3rdvirial2b}
  B_3^0(T) = -\frac{8}{3}\pi^2\int\mathrm{d}R_{12}\mathrm{d}R_{13}\,
  R_{12}^2R_{13}^2 \bigl< f_{12}f_{13}
  f_{23} \sin\vartheta_{(2,1,3)}\bigr>_{\bm{\Omega}_1, \bm{\Omega}_2, 
  \bm{\Omega}_3, \vartheta_{(2,1,3)}},
\end{equation}
\begin{equation}\label{eq:3rdvirial3b}
  \Delta B_3^{3B}(T) = -\frac{8}{3}\pi^2\int\mathrm{d}R_{12}\mathrm{d}R_{13}\,
  R_{12}^2 R_{13}^2
     \biggl< \bigl[\mathrm{e}^{-\beta V_{1,2,3}^{3B}} - 1\bigr]
     \mathrm{e}^{-\beta (V_{1,2}^{2B} + V_{1,3}^{2B} + V_{2,3}^{2B})}
     \sin\vartheta_{(2,1,3)} 
     \biggr>_{\bm{\Omega_1}, \bm{\Omega_2}, \bm{\Omega_3}, \vartheta_{(2,1,3)}}.
\end{equation}
\end{widetext}
Following the work of Tainter et al.,\cite{Tainter2011} we computed the third virial
coefficient by fixing one molecule at the origin, evaluating two
radial integrals through the two-dimensional Simpson rule, and 
using Monte Carlo integration for the 9 orientational degrees of
freedom and the angle $\vartheta_{(2,1,3)}$ between the centers of
mass of molecules 2, 1, 3 at each radial grid point
(using $10^6$ monomer orientations). This integration
strategy was demonstrated in Ref.~\citenum{Tainter2011} to recover
the results from the more efficient Mayer sampling
approach.\cite{Benjamin2007} Our implementation reproduces the
data from Ref.~\citenum{Tainter2011} for the E3B model.

\begin{figure}                
\iffigures
   \includegraphics[width=7.5cm]{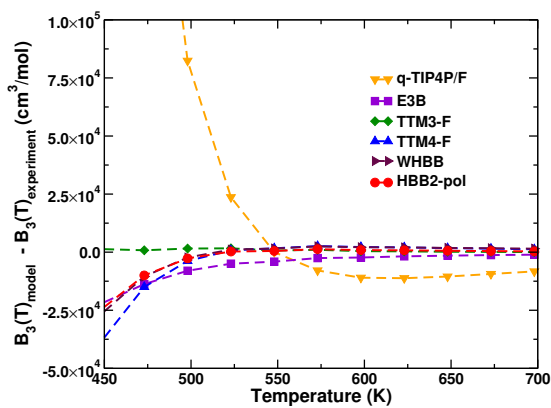}
\fi
   \caption{\label{fig:virial3}Rigid, classical third virial coefficient using vibrationally-averaged 
   monomer geometries. Experimental data from Ref. \citenum{Kell1989}.}
\end{figure}

Rigid, classical third virial coefficients using
vibrationally-averaged monomer geometries are 
reported in figure \ref{fig:virial3}. At lower temperatures, HBB2-pol
agrees with WHBB. In the high temperature limit, however, HBB2-pol compare more favorably
with experiment than WHBB, which is 
consistent with HBB2-pol providing a more accurate description of the 3B interactions.
The errors in the 2B and 3B interactions
of TTM3-F appear to cancel one another, 
resulting in an third virial coefficient that is remarkably
close to experiment. Much work has been 
invested in exploring the role of nuclear quantum effects
\cite{Garberoglio2012,Schenter2002,Bukowski2008} and monomer
flexibility.\cite{Shaul2011}
While the exploration of the monomer configuration on
the second virial coefficient indicates that 
flexibility and nuclear quantum effects are important, these
results are by no means conclusive. We will pursue a more
rigorous characterization of the effect of flexibility and
nuclear quantization in future work.

\section{\label{sec:summary}Summary}
In this study, the accuracy of several force fields, semiempirical methods,
DFT, and \ai-based models in reproducing the two- and three-body
water interactions was assessed against BSSE-corrected CCSD(T)/aug-cc-pVTZ data. 
Our analysis of the many-body expansion of the
interaction energy indicates that defects inherent to polarizable models, which are non-negligable
when molecules are close to one another, can
be effectively corrected through an explicit short-range term
expressed in terms of permutationally invariant polynomials.
Based on these findings, we developed a new water model, HBB2-pol, that
is derived entirely from ``first principles''.
HBB2-pol achieves excellent accuracy with respect to the CCSD(T) data
for the two- and three-body interactions, isomer relative energies of small clusters, 
and second and third virial coefficients.
Importantly, the inclusion of explicit polarization 
in the three-body interaction term enables the use of relatively low-degree polynomials, which,
in turn, results in a significant decrease in the computational cost associated with HBB2-pol 
relative to other \ai-based models.
Through its combined accuracy and computational efficiency, HBB2-pol
thus opens the doorway to fully ``first principles'' simulations of water in the condensed
phases, which will help resolve current controversies\cite{Wernet2004,Clark2010,Pieniazek2011,Nihonyanagi2011,Kumar2008a,Limmer2011}.

\section{Acknowledgement}
This research was supported by the National Science Foundation through grant 
CHE-1111364. We are grateful to the National Science Foundation for a generous 
allocation of computing time on Xsede resources (award TG-CHE110009). 
Additionally, we would like to thank Chris Mundy and Greg Schenter for their 
assistance in calculations involving SCP-NDDO.

%\bibliography{threebody}% Produces the bibliography via BibTeX.

\begin{mcitethebibliography}{92}
\providecommand*\natexlab[1]{#1}
\providecommand*\mciteSetBstSublistMode[1]{}
\providecommand*\mciteSetBstMaxWidthForm[2]{}
\providecommand*\mciteBstWouldAddEndPuncttrue
  {\def\EndOfBibitem{\unskip.}}
\providecommand*\mciteBstWouldAddEndPunctfalse
  {\let\EndOfBibitem\relax}
\providecommand*\mciteSetBstMidEndSepPunct[3]{}
\providecommand*\mciteSetBstSublistLabelBeginEnd[3]{}
\providecommand*\EndOfBibitem{}
\mciteSetBstSublistMode{f}
\mciteSetBstMaxWidthForm{subitem}{(\alph{mcitesubitemcount})}
\mciteSetBstSublistLabelBeginEnd
  {\mcitemaxwidthsubitemform\space}
  {\relax}
  {\relax}

\bibitem[Dahlke et~al.(2008)Dahlke, Olson, Leverentz, and Truhlar]{Dahlke2008}
Dahlke,~E.; Olson,~R.; Leverentz,~H.; Truhlar,~D. \emph{J. Phys. Chem. A}
  \textbf{2008}, \emph{112}, 3976--84\relax
\mciteBstWouldAddEndPuncttrue
\mciteSetBstMidEndSepPunct{\mcitedefaultmidpunct}
{\mcitedefaultendpunct}{\mcitedefaultseppunct}\relax
\EndOfBibitem
\bibitem[Bates and Tschumper(2009)Bates, and Tschumper]{Bates2009a}
Bates,~D.; Tschumper,~G. \emph{J. Phys. Chem. A} \textbf{2009}, \emph{113},
  3555--3559\relax
\mciteBstWouldAddEndPuncttrue
\mciteSetBstMidEndSepPunct{\mcitedefaultmidpunct}
{\mcitedefaultendpunct}{\mcitedefaultseppunct}\relax
\EndOfBibitem
\bibitem[G\'{o}ra et~al.(2011)G\'{o}ra, Podeszwa, Cencek, and
  Szalewicz]{Gora2011}
G\'{o}ra,~U.; Podeszwa,~R.; Cencek,~W.; Szalewicz,~K. \emph{J. Chem. Phys.}
  \textbf{2011}, \emph{135}, 224102\relax
\mciteBstWouldAddEndPuncttrue
\mciteSetBstMidEndSepPunct{\mcitedefaultmidpunct}
{\mcitedefaultendpunct}{\mcitedefaultseppunct}\relax
\EndOfBibitem
\bibitem[Soper and Benmore(2008)Soper, and Benmore]{Soper2008}
Soper,~A.; Benmore,~C. \emph{Phys. Rev. Lett.} \textbf{2008}, \emph{101},
  065502\relax
\mciteBstWouldAddEndPuncttrue
\mciteSetBstMidEndSepPunct{\mcitedefaultmidpunct}
{\mcitedefaultendpunct}{\mcitedefaultseppunct}\relax
\EndOfBibitem
\bibitem[Paesani and Voth(2009)Paesani, and Voth]{Paesani2009}
Paesani,~F.; Voth,~G. \emph{J. Phys. Chem. B} \textbf{2009}, \emph{113},
  5702--19\relax
\mciteBstWouldAddEndPuncttrue
\mciteSetBstMidEndSepPunct{\mcitedefaultmidpunct}
{\mcitedefaultendpunct}{\mcitedefaultseppunct}\relax
\EndOfBibitem
\bibitem[Wang et~al.(2012)Wang, Babin, Bowman, and Paesani]{Wang2012b}
Wang,~Y.; Babin,~V.; Bowman,~J.; Paesani,~F. \emph{J. Am. Chem. Soc.}
  \textbf{2012}, \emph{134}, 11116--9\relax
\mciteBstWouldAddEndPuncttrue
\mciteSetBstMidEndSepPunct{\mcitedefaultmidpunct}
{\mcitedefaultendpunct}{\mcitedefaultseppunct}\relax
\EndOfBibitem
\bibitem[Markland and Berne(2012)Markland, and Berne]{Markland2012}
Markland,~T.; Berne,~B. \emph{Proc. Natl. Acad. Sci.} \textbf{2012},
  \emph{109}, 7988--7991\relax
\mciteBstWouldAddEndPuncttrue
\mciteSetBstMidEndSepPunct{\mcitedefaultmidpunct}
{\mcitedefaultendpunct}{\mcitedefaultseppunct}\relax
\EndOfBibitem
\bibitem[Hankins et~al.(1970)Hankins, Moskowitz, and Stillinger]{Hankins1970}
Hankins,~D.; Moskowitz,~J.; Stillinger,~F. \emph{J. Chem. Phys.} \textbf{1970},
  \emph{53}, 4544--4554\relax
\mciteBstWouldAddEndPuncttrue
\mciteSetBstMidEndSepPunct{\mcitedefaultmidpunct}
{\mcitedefaultendpunct}{\mcitedefaultseppunct}\relax
\EndOfBibitem
\bibitem[Xantheas(1994)]{Xantheas1994}
Xantheas,~S. \emph{J. Chem. Phys.} \textbf{1994}, \emph{100}, 7523--7534\relax
\mciteBstWouldAddEndPuncttrue
\mciteSetBstMidEndSepPunct{\mcitedefaultmidpunct}
{\mcitedefaultendpunct}{\mcitedefaultseppunct}\relax
\EndOfBibitem
\bibitem[Xantheas(2000)]{Xantheas2000}
Xantheas,~S. \emph{Chem. Phys.} \textbf{2000}, \emph{258}, 225--231\relax
\mciteBstWouldAddEndPuncttrue
\mciteSetBstMidEndSepPunct{\mcitedefaultmidpunct}
{\mcitedefaultendpunct}{\mcitedefaultseppunct}\relax
\EndOfBibitem
\bibitem[Defusco et~al.(2007)Defusco, Schofield, and Jordan]{Defusco2007}
Defusco,~A.; Schofield,~D.; Jordan,~K. \emph{Mol. Phys.} \textbf{2007},
  \emph{105}, 2681--2696\relax
\mciteBstWouldAddEndPuncttrue
\mciteSetBstMidEndSepPunct{\mcitedefaultmidpunct}
{\mcitedefaultendpunct}{\mcitedefaultseppunct}\relax
\EndOfBibitem
\bibitem[Kumar et~al.(2010)Kumar, Wang, Jenness, and Jordan]{Kumar2010}
Kumar,~R.; Wang,~F.; Jenness,~G.; Jordan,~K. \emph{J. Chem. Phys.}
  \textbf{2010}, \emph{132}, 014309\relax
\mciteBstWouldAddEndPuncttrue
\mciteSetBstMidEndSepPunct{\mcitedefaultmidpunct}
{\mcitedefaultendpunct}{\mcitedefaultseppunct}\relax
\EndOfBibitem
\bibitem[Hodges et~al.(1997)Hodges, Stone, and Xantheas]{Hodges1997}
Hodges,~M.; Stone,~A.; Xantheas,~S. \emph{J. Phys. Chem. A} \textbf{1997},
  \emph{101}, 9163--9168\relax
\mciteBstWouldAddEndPuncttrue
\mciteSetBstMidEndSepPunct{\mcitedefaultmidpunct}
{\mcitedefaultendpunct}{\mcitedefaultseppunct}\relax
\EndOfBibitem
\bibitem[Ojamie and Hermansson(1994)Ojamie, and Hermansson]{Ojamie1994}
Ojamie,~L.; Hermansson,~K. \emph{J. Phys. Chem.} \textbf{1994}, \emph{98},
  4271--4282\relax
\mciteBstWouldAddEndPuncttrue
\mciteSetBstMidEndSepPunct{\mcitedefaultmidpunct}
{\mcitedefaultendpunct}{\mcitedefaultseppunct}\relax
\EndOfBibitem
\bibitem[Pedulla et~al.(1996)Pedulla, Vila, and Jordan]{Pedulla1996}
Pedulla,~J.; Vila,~F.; Jordan,~K. \emph{J. Chem. Phys.} \textbf{1996},
  \emph{105}, 11091\relax
\mciteBstWouldAddEndPuncttrue
\mciteSetBstMidEndSepPunct{\mcitedefaultmidpunct}
{\mcitedefaultendpunct}{\mcitedefaultseppunct}\relax
\EndOfBibitem
\bibitem[Cui et~al.(2006)Cui, Liu, and Jordan]{Cui2006}
Cui,~J.; Liu,~H.; Jordan,~K. \emph{J. Phys. Chem. B} \textbf{2006}, \emph{110},
  18872--18878\relax
\mciteBstWouldAddEndPuncttrue
\mciteSetBstMidEndSepPunct{\mcitedefaultmidpunct}
{\mcitedefaultendpunct}{\mcitedefaultseppunct}\relax
\EndOfBibitem
\bibitem[Hermann et~al.(2007)Hermann, Krawczyk, Lein, Schwerdtfeger, Hamilton,
  and Stewart]{Hermann2007}
Hermann,~A.; Krawczyk,~R.; Lein,~M.; Schwerdtfeger,~P.; Hamilton,~I.;
  Stewart,~J. \emph{Phys. Rev. A} \textbf{2007}, \emph{76}, 013202\relax
\mciteBstWouldAddEndPuncttrue
\mciteSetBstMidEndSepPunct{\mcitedefaultmidpunct}
{\mcitedefaultendpunct}{\mcitedefaultseppunct}\relax
\EndOfBibitem
\bibitem[Jorgensen et~al.(1983)Jorgensen, Chandrasekhar, Madura, Impey, and
  Klein]{Jorgensen1983}
Jorgensen,~W.; Chandrasekhar,~J.; Madura,~J.; Impey,~R.; Klein,~M. \emph{J.
  Chem. Phys.} \textbf{1983}, \emph{79}, 926\relax
\mciteBstWouldAddEndPuncttrue
\mciteSetBstMidEndSepPunct{\mcitedefaultmidpunct}
{\mcitedefaultendpunct}{\mcitedefaultseppunct}\relax
\EndOfBibitem
\bibitem[Berendsen et~al.(1981)Berendsen, Postma, van Gunsteren, and
  Hermans]{Berendsen1981}
Berendsen,~H.; Postma,~J.; van Gunsteren,~W.; Hermans,~J. In
  \emph{Intermolecular Forces}; Pullman,~B., Ed.; Reidel Publishing Company:
  Dordrecht, 1981; pp 333--342\relax
\mciteBstWouldAddEndPuncttrue
\mciteSetBstMidEndSepPunct{\mcitedefaultmidpunct}
{\mcitedefaultendpunct}{\mcitedefaultseppunct}\relax
\EndOfBibitem
\bibitem[Habershon et~al.(2009)Habershon, Markland, and
  Manolopoulos]{Habershon2009}
Habershon,~S.; Markland,~T.; Manolopoulos,~D. \emph{J. Chem. Phys.}
  \textbf{2009}, \emph{131}, 024501\relax
\mciteBstWouldAddEndPuncttrue
\mciteSetBstMidEndSepPunct{\mcitedefaultmidpunct}
{\mcitedefaultendpunct}{\mcitedefaultseppunct}\relax
\EndOfBibitem
\bibitem[Paesani et~al.(2006)Paesani, Zhang, Case, Cheatham, and
  Voth]{Paesani2006}
Paesani,~F.; Zhang,~W.; Case,~D.; Cheatham,~T.; Voth,~G. \emph{J. Chem. Phys.}
  \textbf{2006}, \emph{125}, 184507\relax
\mciteBstWouldAddEndPuncttrue
\mciteSetBstMidEndSepPunct{\mcitedefaultmidpunct}
{\mcitedefaultendpunct}{\mcitedefaultseppunct}\relax
\EndOfBibitem
\bibitem[Vega and Abascal(2011)Vega, and Abascal]{Vega2011}
Vega,~C.; Abascal,~J. \emph{Phys. Chem. Chem. Phys.} \textbf{2011}, \emph{13},
  19663--19688\relax
\mciteBstWouldAddEndPuncttrue
\mciteSetBstMidEndSepPunct{\mcitedefaultmidpunct}
{\mcitedefaultendpunct}{\mcitedefaultseppunct}\relax
\EndOfBibitem
\bibitem[Kumar and Skinner(2008)Kumar, and Skinner]{Kumar2008}
Kumar,~R.; Skinner,~J. \emph{J. Phys. Chem. B} \textbf{2008}, \emph{112},
  8311--8318\relax
\mciteBstWouldAddEndPuncttrue
\mciteSetBstMidEndSepPunct{\mcitedefaultmidpunct}
{\mcitedefaultendpunct}{\mcitedefaultseppunct}\relax
\EndOfBibitem
\bibitem[Tainter et~al.(2011)Tainter, Pieniazek, Lin, and Skinner]{Tainter2011}
Tainter,~C.; Pieniazek,~P.; Lin,~Y.; Skinner,~J. \emph{J. Chem. Phys.}
  \textbf{2011}, \emph{134}, 184501\relax
\mciteBstWouldAddEndPuncttrue
\mciteSetBstMidEndSepPunct{\mcitedefaultmidpunct}
{\mcitedefaultendpunct}{\mcitedefaultseppunct}\relax
\EndOfBibitem
\bibitem[Tainter and Skinner(2012)Tainter, and Skinner]{Tainter2012}
Tainter,~C.; Skinner,~J. \emph{J. Chem. Phys.} \textbf{2012}, \emph{137},
  104304\relax
\mciteBstWouldAddEndPuncttrue
\mciteSetBstMidEndSepPunct{\mcitedefaultmidpunct}
{\mcitedefaultendpunct}{\mcitedefaultseppunct}\relax
\EndOfBibitem
\bibitem[Lopes et~al.(2009)Lopes, Roux, and Mackerell]{Lopes2009}
Lopes,~P.; Roux,~B.; Mackerell,~A. \emph{Theor. Chem. Acc.} \textbf{2009},
  \emph{124}, 11--28\relax
\mciteBstWouldAddEndPuncttrue
\mciteSetBstMidEndSepPunct{\mcitedefaultmidpunct}
{\mcitedefaultendpunct}{\mcitedefaultseppunct}\relax
\EndOfBibitem
\bibitem[Applequist et~al.(1972)Applequist, Carl, and Fung]{Applequist1972}
Applequist,~J.; Carl,~J.; Fung,~K. \emph{J. Am. Chem. Soc.} \textbf{1972},
  \emph{94}, 2952--2960\relax
\mciteBstWouldAddEndPuncttrue
\mciteSetBstMidEndSepPunct{\mcitedefaultmidpunct}
{\mcitedefaultendpunct}{\mcitedefaultseppunct}\relax
\EndOfBibitem
\bibitem[Thole(1981)]{Thole1981}
Thole,~B. \emph{Chem. Phys.} \textbf{1981}, \emph{59}, 341--350\relax
\mciteBstWouldAddEndPuncttrue
\mciteSetBstMidEndSepPunct{\mcitedefaultmidpunct}
{\mcitedefaultendpunct}{\mcitedefaultseppunct}\relax
\EndOfBibitem
\bibitem[Fanourgakis and Xantheas(2008)Fanourgakis, and
  Xantheas]{Fanourgakis2008a}
Fanourgakis,~G.; Xantheas,~S. \emph{J. Chem. Phys.} \textbf{2008}, \emph{128},
  074506\relax
\mciteBstWouldAddEndPuncttrue
\mciteSetBstMidEndSepPunct{\mcitedefaultmidpunct}
{\mcitedefaultendpunct}{\mcitedefaultseppunct}\relax
\EndOfBibitem
\bibitem[Burnham et~al.(2008)Burnham, Anick, Mankoo, and Reiter]{Burnham2008}
Burnham,~C.; Anick,~D.; Mankoo,~P.; Reiter,~G. \emph{J. Chem. Phys.}
  \textbf{2008}, \emph{128}, 154519\relax
\mciteBstWouldAddEndPuncttrue
\mciteSetBstMidEndSepPunct{\mcitedefaultmidpunct}
{\mcitedefaultendpunct}{\mcitedefaultseppunct}\relax
\EndOfBibitem
\bibitem[Ren and Ponder(2003)Ren, and Ponder]{Ren2003}
Ren,~P.; Ponder,~J. \emph{J. Phys. Chem. B} \textbf{2003}, \emph{107},
  5933--5947\relax
\mciteBstWouldAddEndPuncttrue
\mciteSetBstMidEndSepPunct{\mcitedefaultmidpunct}
{\mcitedefaultendpunct}{\mcitedefaultseppunct}\relax
\EndOfBibitem
\bibitem[Stewart(1989)]{Stewart1989}
Stewart,~J. \emph{J. Comput. Chem.} \textbf{1989}, \emph{10}, 209--220\relax
\mciteBstWouldAddEndPuncttrue
\mciteSetBstMidEndSepPunct{\mcitedefaultmidpunct}
{\mcitedefaultendpunct}{\mcitedefaultseppunct}\relax
\EndOfBibitem
\bibitem[Bernal-Uruchurtu and Ruiz-L\'{o}pez(2000)Bernal-Uruchurtu, and
  Ruiz-L\'{o}pez]{Bernal-Uruchurtu2000a}
Bernal-Uruchurtu,~M.; Ruiz-L\'{o}pez,~M. \emph{Chem. Phys. Lett.}
  \textbf{2000}, \emph{330}, 118--124\relax
\mciteBstWouldAddEndPuncttrue
\mciteSetBstMidEndSepPunct{\mcitedefaultmidpunct}
{\mcitedefaultendpunct}{\mcitedefaultseppunct}\relax
\EndOfBibitem
\bibitem[Chang et~al.(2008)Chang, Schenter, and Garrett]{Chang2008}
Chang,~D.; Schenter,~G.; Garrett,~B. \emph{J. Chem. Phys.} \textbf{2008},
  \emph{128}, 164111\relax
\mciteBstWouldAddEndPuncttrue
\mciteSetBstMidEndSepPunct{\mcitedefaultmidpunct}
{\mcitedefaultendpunct}{\mcitedefaultseppunct}\relax
\EndOfBibitem
\bibitem[Murdachaew et~al.(2011)Murdachaew, Mundy, Schenter, Laino, and
  Hutter]{Murdachaew2011a}
Murdachaew,~G.; Mundy,~C.; Schenter,~G.; Laino,~T.; Hutter,~J. \emph{J. Phys.
  Chem. A} \textbf{2011}, \emph{115}, 6046--53\relax
\mciteBstWouldAddEndPuncttrue
\mciteSetBstMidEndSepPunct{\mcitedefaultmidpunct}
{\mcitedefaultendpunct}{\mcitedefaultseppunct}\relax
\EndOfBibitem
\bibitem[Becke(1988)]{Becke1988}
Becke,~A. \emph{Phys. Rev. A} \textbf{1988}, \emph{38}, 3098--3100\relax
\mciteBstWouldAddEndPuncttrue
\mciteSetBstMidEndSepPunct{\mcitedefaultmidpunct}
{\mcitedefaultendpunct}{\mcitedefaultseppunct}\relax
\EndOfBibitem
\bibitem[Lee et~al.(1988)Lee, Yang, and Parr]{Lee1988}
Lee,~C.; Yang,~W.; Parr,~R. \emph{Phys. Rev. B} \textbf{1988}, \emph{37},
  785--789\relax
\mciteBstWouldAddEndPuncttrue
\mciteSetBstMidEndSepPunct{\mcitedefaultmidpunct}
{\mcitedefaultendpunct}{\mcitedefaultseppunct}\relax
\EndOfBibitem
\bibitem[Perdew et~al.(1996)Perdew, Burke, and Ernzerhof]{Perdew1996}
Perdew,~J.; Burke,~K.; Ernzerhof,~M. \emph{Phys. Rev. Lett.} \textbf{1996},
  \emph{77}, 3865--3868\relax
\mciteBstWouldAddEndPuncttrue
\mciteSetBstMidEndSepPunct{\mcitedefaultmidpunct}
{\mcitedefaultendpunct}{\mcitedefaultseppunct}\relax
\EndOfBibitem
\bibitem[Grimme(2004)]{Grimme2004}
Grimme,~S. \emph{J. Comput. Chem.} \textbf{2004}, \emph{25}, 1463--1473\relax
\mciteBstWouldAddEndPuncttrue
\mciteSetBstMidEndSepPunct{\mcitedefaultmidpunct}
{\mcitedefaultendpunct}{\mcitedefaultseppunct}\relax
\EndOfBibitem
\bibitem[Grimme(2006)]{Grimme2006}
Grimme,~S. \emph{J. Comput. Chem.} \textbf{2006}, \emph{27}, 1787--1799\relax
\mciteBstWouldAddEndPuncttrue
\mciteSetBstMidEndSepPunct{\mcitedefaultmidpunct}
{\mcitedefaultendpunct}{\mcitedefaultseppunct}\relax
\EndOfBibitem
\bibitem[Fulton et~al.(2010)Fulton, Schenter, Baer, Mundy, Dang, and
  Balasubramanian]{Fulton2010}
Fulton,~J.; Schenter,~G.; Baer,~M.; Mundy,~C.; Dang,~L.; Balasubramanian,~M.
  \emph{J. Phys. Chem. B} \textbf{2010}, \emph{114}, 12926--12937\relax
\mciteBstWouldAddEndPuncttrue
\mciteSetBstMidEndSepPunct{\mcitedefaultmidpunct}
{\mcitedefaultendpunct}{\mcitedefaultseppunct}\relax
\EndOfBibitem
\bibitem[Ma et~al.(2012)Ma, Zhang, and Tuckerman]{Ma2012}
Ma,~Z.; Zhang,~Y.; Tuckerman,~M. \emph{J. Chem. Phys.} \textbf{2012},
  \emph{044506}, 044506\relax
\mciteBstWouldAddEndPuncttrue
\mciteSetBstMidEndSepPunct{\mcitedefaultmidpunct}
{\mcitedefaultendpunct}{\mcitedefaultseppunct}\relax
\EndOfBibitem
\bibitem[Grimme et~al.(2010)Grimme, Antony, Ehrlich, and Krieg]{Grimme2010}
Grimme,~S.; Antony,~J.; Ehrlich,~S.; Krieg,~H. \emph{J. Chem. Phys.}
  \textbf{2010}, \emph{132}, 154104\relax
\mciteBstWouldAddEndPuncttrue
\mciteSetBstMidEndSepPunct{\mcitedefaultmidpunct}
{\mcitedefaultendpunct}{\mcitedefaultseppunct}\relax
\EndOfBibitem
\bibitem[Dion et~al.(2004)Dion, Rydberg, Schr\"{o}der, Langreth, and
  Lundqvist]{Dion2004}
Dion,~M.; Rydberg,~H.; Schr\"{o}der,~E.; Langreth,~D.; Lundqvist,~B.
  \emph{Phys. Rev. Lett} \textbf{2004}, \emph{92}, 246401\relax
\mciteBstWouldAddEndPuncttrue
\mciteSetBstMidEndSepPunct{\mcitedefaultmidpunct}
{\mcitedefaultendpunct}{\mcitedefaultseppunct}\relax
\EndOfBibitem
\bibitem[Lee et~al.(2010)Lee, Murray, Kong, Lundqvist, and Langreth]{Lee2010}
Lee,~K.; Murray,~E.; Kong,~L.; Lundqvist,~B.; Langreth,~D. \emph{Phys. Rev. B}
  \textbf{2010}, \emph{82}, 081101\relax
\mciteBstWouldAddEndPuncttrue
\mciteSetBstMidEndSepPunct{\mcitedefaultmidpunct}
{\mcitedefaultendpunct}{\mcitedefaultseppunct}\relax
\EndOfBibitem
\bibitem[Vydrov and {Van Voorhis}(2010)Vydrov, and {Van Voorhis}]{Vydrov2010}
Vydrov,~O.; {Van Voorhis},~T. \emph{J. Chem. Phys.} \textbf{2010}, \emph{133},
  244103\relax
\mciteBstWouldAddEndPuncttrue
\mciteSetBstMidEndSepPunct{\mcitedefaultmidpunct}
{\mcitedefaultendpunct}{\mcitedefaultseppunct}\relax
\EndOfBibitem
\bibitem[Wang et~al.(2011)Wang, Rom\'{a}n-P\'{e}rez, Soler, Artacho, and
  Fern\'{a}ndez-Serra]{Wang2011e}
Wang,~J.; Rom\'{a}n-P\'{e}rez,~G.; Soler,~J.; Artacho,~E.;
  Fern\'{a}ndez-Serra,~M. \emph{J. Chem. Phys.} \textbf{2011}, \emph{134},
  024516\relax
\mciteBstWouldAddEndPuncttrue
\mciteSetBstMidEndSepPunct{\mcitedefaultmidpunct}
{\mcitedefaultendpunct}{\mcitedefaultseppunct}\relax
\EndOfBibitem
\bibitem[Murray and Galli(2012)Murray, and Galli]{Murray2012}
Murray,~E.; Galli,~G. \emph{Phys. Rev. Lett.} \textbf{2012}, \emph{108},
  105502\relax
\mciteBstWouldAddEndPuncttrue
\mciteSetBstMidEndSepPunct{\mcitedefaultmidpunct}
{\mcitedefaultendpunct}{\mcitedefaultseppunct}\relax
\EndOfBibitem
\bibitem[Bukowski et~al.(2008)Bukowski, Szalewicz, Groenenboom, and {van Der
  Avoird}]{Bukowski2008}
Bukowski,~R.; Szalewicz,~K.; Groenenboom,~G.; {van Der Avoird},~A. \emph{J.
  Chem. Phys.} \textbf{2008}, \emph{128}, 094314\relax
\mciteBstWouldAddEndPuncttrue
\mciteSetBstMidEndSepPunct{\mcitedefaultmidpunct}
{\mcitedefaultendpunct}{\mcitedefaultseppunct}\relax
\EndOfBibitem
\bibitem[Bukowski et~al.(2008)Bukowski, Szalewicz, Groenenboom, and {van Der
  Avoird}]{Bukowski2008a}
Bukowski,~R.; Szalewicz,~K.; Groenenboom,~G.; {van Der Avoird},~A. \emph{J.
  Chem. Phys.} \textbf{2008}, \emph{128}, 094313\relax
\mciteBstWouldAddEndPuncttrue
\mciteSetBstMidEndSepPunct{\mcitedefaultmidpunct}
{\mcitedefaultendpunct}{\mcitedefaultseppunct}\relax
\EndOfBibitem
\bibitem[Wang et~al.(2011)Wang, Huang, Shepler, Braams, and Bowman]{Wang2011b}
Wang,~Y.; Huang,~X.; Shepler,~B.; Braams,~B.; Bowman,~J. \emph{J. Chem. Phys.}
  \textbf{2011}, \emph{134}, 094509\relax
\mciteBstWouldAddEndPuncttrue
\mciteSetBstMidEndSepPunct{\mcitedefaultmidpunct}
{\mcitedefaultendpunct}{\mcitedefaultseppunct}\relax
\EndOfBibitem
\bibitem[Bukowski et~al.(2007)Bukowski, Szalewicz, Groenenboom, and van~der
  Avoird]{Bukowski2007}
Bukowski,~R.; Szalewicz,~K.; Groenenboom,~G.; van~der Avoird,~A. \emph{Science}
  \textbf{2007}, \emph{315}, 1249--52\relax
\mciteBstWouldAddEndPuncttrue
\mciteSetBstMidEndSepPunct{\mcitedefaultmidpunct}
{\mcitedefaultendpunct}{\mcitedefaultseppunct}\relax
\EndOfBibitem
\bibitem[Leforestier et~al.(2012)Leforestier, Szalewicz, and van~der
  Avoird]{Leforestier2012}
Leforestier,~C.; Szalewicz,~K.; van~der Avoird,~A. \emph{J. Chem. Phys.}
  \textbf{2012}, \emph{137}, 014305\relax
\mciteBstWouldAddEndPuncttrue
\mciteSetBstMidEndSepPunct{\mcitedefaultmidpunct}
{\mcitedefaultendpunct}{\mcitedefaultseppunct}\relax
\EndOfBibitem
\bibitem[Raghavachari et~al.(1989)Raghavachari, Truck, Pople, and
  Head-Gordon]{Raghavachari1989}
Raghavachari,~K.; Truck,~G.; Pople,~J.; Head-Gordon,~M. \emph{Chem. Phys.
  Lett.} \textbf{1989}, \emph{157}, 479--483\relax
\mciteBstWouldAddEndPuncttrue
\mciteSetBstMidEndSepPunct{\mcitedefaultmidpunct}
{\mcitedefaultendpunct}{\mcitedefaultseppunct}\relax
\EndOfBibitem
\bibitem[Dunning(1989)]{Dunning1989}
Dunning,~T. \emph{J. Chem. Phys.} \textbf{1989}, \emph{90}, 1007\relax
\mciteBstWouldAddEndPuncttrue
\mciteSetBstMidEndSepPunct{\mcitedefaultmidpunct}
{\mcitedefaultendpunct}{\mcitedefaultseppunct}\relax
\EndOfBibitem
\bibitem[Boys and Bernardi(1970)Boys, and Bernardi]{Boys1970}
Boys,~S.; Bernardi,~F. \emph{Mol. Phys.} \textbf{1970}, \emph{19}, 553\relax
\mciteBstWouldAddEndPuncttrue
\mciteSetBstMidEndSepPunct{\mcitedefaultmidpunct}
{\mcitedefaultendpunct}{\mcitedefaultseppunct}\relax
\EndOfBibitem
\bibitem[Weigend and Ahlrichs(2005)Weigend, and Ahlrichs]{Weigend2005}
Weigend,~F.; Ahlrichs,~R. \emph{Phys. Chem. Chem. Phys.} \textbf{2005},
  \emph{7}, 3297--3305\relax
\mciteBstWouldAddEndPuncttrue
\mciteSetBstMidEndSepPunct{\mcitedefaultmidpunct}
{\mcitedefaultendpunct}{\mcitedefaultseppunct}\relax
\EndOfBibitem
\bibitem[Schafer et~al.(1994)Schafer, Huber, and Ahlrichs]{Schafer1994}
Schafer,~A.; Huber,~C.; Ahlrichs,~R. \emph{J. Chem. Phys.} \textbf{1994},
  \emph{100}, 5829--5835\relax
\mciteBstWouldAddEndPuncttrue
\mciteSetBstMidEndSepPunct{\mcitedefaultmidpunct}
{\mcitedefaultendpunct}{\mcitedefaultseppunct}\relax
\EndOfBibitem
\bibitem[Wennmohs and Neese(2008)Wennmohs, and Neese]{Wennmohs2008}
Wennmohs,~F.; Neese,~F. \emph{Chem. Phys.} \textbf{2008}, \emph{343},
  217--230\relax
\mciteBstWouldAddEndPuncttrue
\mciteSetBstMidEndSepPunct{\mcitedefaultmidpunct}
{\mcitedefaultendpunct}{\mcitedefaultseppunct}\relax
\EndOfBibitem
\bibitem[Walker et~al.(2008)Walker, Crowley, and Case]{Walker2008}
Walker,~R.~C.; Crowley,~M.~F.; Case,~D.~A. \emph{J. Comput. Chem.}
  \textbf{2008}, \emph{29}, 1019--1031\relax
\mciteBstWouldAddEndPuncttrue
\mciteSetBstMidEndSepPunct{\mcitedefaultmidpunct}
{\mcitedefaultendpunct}{\mcitedefaultseppunct}\relax
\EndOfBibitem
\bibitem[cp2()]{cp2k}
\emph{http://cp2k.berlios.de/} 2000--2012\relax
\mciteBstWouldAddEndPuncttrue
\mciteSetBstMidEndSepPunct{\mcitedefaultmidpunct}
{\mcitedefaultendpunct}{\mcitedefaultseppunct}\relax
\EndOfBibitem
\bibitem[Murdachaew et~al.(2011)Murdachaew, Mundy, Schenter, Laino, and
  Hutter]{Murdachaew2011}
Murdachaew,~G.; Mundy,~C.; Schenter,~G.; Laino,~T.; Hutter,~J. \emph{J. Phys.
  Chem. A} \textbf{2011}, \emph{115}, 6046--6053\relax
\mciteBstWouldAddEndPuncttrue
\mciteSetBstMidEndSepPunct{\mcitedefaultmidpunct}
{\mcitedefaultendpunct}{\mcitedefaultseppunct}\relax
\EndOfBibitem
\bibitem[Burnham and Xantheas(2002)Burnham, and Xantheas]{Burnham2002c}
Burnham,~C.; Xantheas,~S. \emph{J. Chem. Phys.} \textbf{2002}, \emph{116},
  5115\relax
\mciteBstWouldAddEndPuncttrue
\mciteSetBstMidEndSepPunct{\mcitedefaultmidpunct}
{\mcitedefaultendpunct}{\mcitedefaultseppunct}\relax
\EndOfBibitem
\bibitem[VandeVondele and Hutter(2007)VandeVondele, and
  Hutter]{VandeVondele2007}
VandeVondele,~J.; Hutter,~J. \emph{J. Chem. Phys.} \textbf{2007}, \emph{127},
  114105\relax
\mciteBstWouldAddEndPuncttrue
\mciteSetBstMidEndSepPunct{\mcitedefaultmidpunct}
{\mcitedefaultendpunct}{\mcitedefaultseppunct}\relax
\EndOfBibitem
\bibitem[Becke(1993)]{Becke1993}
Becke,~A. \emph{J. Chem. Phys} \textbf{1993}, \emph{98}, 5648\relax
\mciteBstWouldAddEndPuncttrue
\mciteSetBstMidEndSepPunct{\mcitedefaultmidpunct}
{\mcitedefaultendpunct}{\mcitedefaultseppunct}\relax
\EndOfBibitem
\bibitem[Stephens et~al.(1994)Stephens, Devlin, Chabalowski, and
  Frisch]{Stephens1994}
Stephens,~P.; Devlin,~F.; Chabalowski,~C.; Frisch,~M. \emph{J. Phys. Chem.}
  \textbf{1994}, \emph{98}, 11623--11627\relax
\mciteBstWouldAddEndPuncttrue
\mciteSetBstMidEndSepPunct{\mcitedefaultmidpunct}
{\mcitedefaultendpunct}{\mcitedefaultseppunct}\relax
\EndOfBibitem
\bibitem[Klimes and Michaelides(2012)Klimes, and Michaelides]{Klimes2012}
Klimes,~J.; Michaelides,~A. \emph{J. Chem. Phys.} \textbf{2012}, \emph{137},
  120901\relax
\mciteBstWouldAddEndPuncttrue
\mciteSetBstMidEndSepPunct{\mcitedefaultmidpunct}
{\mcitedefaultendpunct}{\mcitedefaultseppunct}\relax
\EndOfBibitem
\bibitem[Chen and Gordon(1996)Chen, and Gordon]{Chen1996}
Chen,~W.; Gordon,~M. \emph{J. Phys. Chem.} \textbf{1996}, \emph{100},
  14316--14328\relax
\mciteBstWouldAddEndPuncttrue
\mciteSetBstMidEndSepPunct{\mcitedefaultmidpunct}
{\mcitedefaultendpunct}{\mcitedefaultseppunct}\relax
\EndOfBibitem
\bibitem[Shank et~al.(2009)Shank, Wang, Kaledin, Braams, and Bowman]{Shank2009}
Shank,~A.; Wang,~Y.; Kaledin,~A.; Braams,~B.; Bowman,~J. \emph{J. Chem. Phys.}
  \textbf{2009}, \emph{130}, 144314\relax
\mciteBstWouldAddEndPuncttrue
\mciteSetBstMidEndSepPunct{\mcitedefaultmidpunct}
{\mcitedefaultendpunct}{\mcitedefaultseppunct}\relax
\EndOfBibitem
\bibitem[Mas et~al.(2003)Mas, Bukowski, and Szalewicz]{Mas2003}
Mas,~E.; Bukowski,~R.; Szalewicz,~K. \emph{J. Chem. Phys.} \textbf{2003},
  \emph{118}, 4386--4403\relax
\mciteBstWouldAddEndPuncttrue
\mciteSetBstMidEndSepPunct{\mcitedefaultmidpunct}
{\mcitedefaultendpunct}{\mcitedefaultseppunct}\relax
\EndOfBibitem
\bibitem[Caldwell et~al.(1990)Caldwell, Dang, and Kollman]{Caldwell1990}
Caldwell,~J.; Dang,~L.; Kollman,~P. \emph{J. Am. Chem. Soc} \textbf{1990},
  \emph{112}, 9144--9147\relax
\mciteBstWouldAddEndPuncttrue
\mciteSetBstMidEndSepPunct{\mcitedefaultmidpunct}
{\mcitedefaultendpunct}{\mcitedefaultseppunct}\relax
\EndOfBibitem
\bibitem[Galassi(2009)]{Galassi2009}
Galassi,~M. \emph{{GNU Scientific Library Reference Manual}}, 3rd ed.; Network
  Theory Ltd., 2009\relax
\mciteBstWouldAddEndPuncttrue
\mciteSetBstMidEndSepPunct{\mcitedefaultmidpunct}
{\mcitedefaultendpunct}{\mcitedefaultseppunct}\relax
\EndOfBibitem
\bibitem[{Maple 11}()]{maple}
{Maple 11}, \emph{Maplesoft, a division of Waterloo Maple Inc., Waterloo,
  Ontario} \relax
\mciteBstWouldAddEndPunctfalse
\mciteSetBstMidEndSepPunct{\mcitedefaultmidpunct}
{}{\mcitedefaultseppunct}\relax
\EndOfBibitem
\bibitem[Partridge and Schwenke(1997)Partridge, and Schwenke]{Partridge1997a}
Partridge,~H.; Schwenke,~D. \emph{J. Chem. Phys.} \textbf{1997}, \emph{106},
  4618\relax
\mciteBstWouldAddEndPuncttrue
\mciteSetBstMidEndSepPunct{\mcitedefaultmidpunct}
{\mcitedefaultendpunct}{\mcitedefaultseppunct}\relax
\EndOfBibitem
\bibitem[Anderson et~al.(2004)Anderson, Crager, Fedoroff, and
  Tschumper]{Anderson2004}
Anderson,~J.; Crager,~K.; Fedoroff,~L.; Tschumper,~G. \emph{J. Chem. Phys.}
  \textbf{2004}, \emph{121}, 11023--11029\relax
\mciteBstWouldAddEndPuncttrue
\mciteSetBstMidEndSepPunct{\mcitedefaultmidpunct}
{\mcitedefaultendpunct}{\mcitedefaultseppunct}\relax
\EndOfBibitem
\bibitem[Hill(1986)]{Hill1986}
Hill,~T. \emph{{An Introduction to Statistical Thermodynamics}}; Dover,
  1986\relax
\mciteBstWouldAddEndPuncttrue
\mciteSetBstMidEndSepPunct{\mcitedefaultmidpunct}
{\mcitedefaultendpunct}{\mcitedefaultseppunct}\relax
\EndOfBibitem
\bibitem[Mayer and Mayer(1940)Mayer, and Mayer]{Mayer1940}
Mayer,~J.; Mayer,~M. \emph{{Statistical Mechanics}}; John Wiley \& Sons Inc,
  1940\relax
\mciteBstWouldAddEndPuncttrue
\mciteSetBstMidEndSepPunct{\mcitedefaultmidpunct}
{\mcitedefaultendpunct}{\mcitedefaultseppunct}\relax
\EndOfBibitem
\bibitem[Mason and Spurling(1969)Mason, and Spurling]{Mason1969}
Mason,~E.; Spurling,~T. \emph{International encyclopedia of physical chemistry
  and chemical physics. Topic 10, Fluid state, V. 2}; Pergamon Press: New York,
  1969\relax
\mciteBstWouldAddEndPuncttrue
\mciteSetBstMidEndSepPunct{\mcitedefaultmidpunct}
{\mcitedefaultendpunct}{\mcitedefaultseppunct}\relax
\EndOfBibitem
\bibitem[Harvey and Lemmon(2004)Harvey, and Lemmon]{Harvey2004}
Harvey,~A.; Lemmon,~E. \emph{J. Phys. Chem. Ref. Data.} \textbf{2004},
  \emph{33}, 369--376\relax
\mciteBstWouldAddEndPuncttrue
\mciteSetBstMidEndSepPunct{\mcitedefaultmidpunct}
{\mcitedefaultendpunct}{\mcitedefaultseppunct}\relax
\EndOfBibitem
\bibitem[Mas and Szalewicz(1996)Mas, and Szalewicz]{Mas1996}
Mas,~E.; Szalewicz,~K. \emph{J. Chem. Phys.} \textbf{1996}, \emph{104},
  7606\relax
\mciteBstWouldAddEndPuncttrue
\mciteSetBstMidEndSepPunct{\mcitedefaultmidpunct}
{\mcitedefaultendpunct}{\mcitedefaultseppunct}\relax
\EndOfBibitem
\bibitem[Benjamin et~al.(2007)Benjamin, Singh, Schultz, and
  Kofke]{Benjamin2007}
Benjamin,~K.; Singh,~J.; Schultz,~A.; Kofke,~D. \emph{J. Phys. Chem. B}
  \textbf{2007}, \emph{111}, 11463--11473\relax
\mciteBstWouldAddEndPuncttrue
\mciteSetBstMidEndSepPunct{\mcitedefaultmidpunct}
{\mcitedefaultendpunct}{\mcitedefaultseppunct}\relax
\EndOfBibitem
\bibitem[Kell et~al.(1989)Kell, McLaurin, and Whalley]{Kell1989}
Kell,~G.; McLaurin,~G.; Whalley,~E. \emph{Proc. R. Soc. Lond. A} \textbf{1989},
  \emph{425}, 49--71\relax
\mciteBstWouldAddEndPuncttrue
\mciteSetBstMidEndSepPunct{\mcitedefaultmidpunct}
{\mcitedefaultendpunct}{\mcitedefaultseppunct}\relax
\EndOfBibitem
\bibitem[Garberoglio(2012)]{Garberoglio2012}
Garberoglio,~G. \emph{Chem. Phys. Lett.} \textbf{2012}, \emph{525-526},
  19--23\relax
\mciteBstWouldAddEndPuncttrue
\mciteSetBstMidEndSepPunct{\mcitedefaultmidpunct}
{\mcitedefaultendpunct}{\mcitedefaultseppunct}\relax
\EndOfBibitem
\bibitem[Schenter(2002)]{Schenter2002}
Schenter,~G. \emph{J. Chem. Phys.} \textbf{2002}, \emph{117}, 6573\relax
\mciteBstWouldAddEndPuncttrue
\mciteSetBstMidEndSepPunct{\mcitedefaultmidpunct}
{\mcitedefaultendpunct}{\mcitedefaultseppunct}\relax
\EndOfBibitem
\bibitem[Shaul et~al.(2011)Shaul, Schultz, and Kofke]{Shaul2011}
Shaul,~K.; Schultz,~A.; Kofke,~D. \emph{J. Chem. Phys.} \textbf{2011},
  \emph{135}, 124101\relax
\mciteBstWouldAddEndPuncttrue
\mciteSetBstMidEndSepPunct{\mcitedefaultmidpunct}
{\mcitedefaultendpunct}{\mcitedefaultseppunct}\relax
\EndOfBibitem
\bibitem[Wernet et~al.(2004)Wernet, Nordlund, Bergmann, Cavalleri, Odelius,
  Ogasawara, N\"{a}slund, Hirsch, Ojam\"{a}e, Glatzel, Pettersson, and
  Nilsson]{Wernet2004}
Wernet,~P.; Nordlund,~D.; Bergmann,~U.; Cavalleri,~M.; Odelius,~M.;
  Ogasawara,~H.; N\"{a}slund,~L.; Hirsch,~T.; Ojam\"{a}e,~L.; Glatzel,~P.;
  Pettersson,~L.; Nilsson,~A. \emph{Science} \textbf{2004}, \emph{304},
  995--999\relax
\mciteBstWouldAddEndPuncttrue
\mciteSetBstMidEndSepPunct{\mcitedefaultmidpunct}
{\mcitedefaultendpunct}{\mcitedefaultseppunct}\relax
\EndOfBibitem
\bibitem[Clark et~al.(2010)Clark, Cappa, Smith, Saykally, and
  Head-Gordon]{Clark2010}
Clark,~G.; Cappa,~C.; Smith,~J.; Saykally,~R.; Head-Gordon,~T. \emph{Mol.
  Phys.} \textbf{2010}, \emph{108}, 1415--1433\relax
\mciteBstWouldAddEndPuncttrue
\mciteSetBstMidEndSepPunct{\mcitedefaultmidpunct}
{\mcitedefaultendpunct}{\mcitedefaultseppunct}\relax
\EndOfBibitem
\bibitem[Pieniazek et~al.(2011)Pieniazek, Tainter, and Skinner]{Pieniazek2011}
Pieniazek,~P.; Tainter,~C.; Skinner,~J. \emph{J. Am. Chem. Soc.} \textbf{2011},
  \emph{133}, 10360\relax
\mciteBstWouldAddEndPuncttrue
\mciteSetBstMidEndSepPunct{\mcitedefaultmidpunct}
{\mcitedefaultendpunct}{\mcitedefaultseppunct}\relax
\EndOfBibitem
\bibitem[Nihonyanagi et~al.(2011)Nihonyanagi, Ishiyama, Lee, Yamaguchi, Bonn,
  Morita, and Tahara]{Nihonyanagi2011}
Nihonyanagi,~S.; Ishiyama,~T.; Lee,~T.; Yamaguchi,~S.; Bonn,~M.; Morita,~A.;
  Tahara,~T. \emph{J. Am. Chem. Soc.} \textbf{2011}, \emph{133},
  16875--16880\relax
\mciteBstWouldAddEndPuncttrue
\mciteSetBstMidEndSepPunct{\mcitedefaultmidpunct}
{\mcitedefaultendpunct}{\mcitedefaultseppunct}\relax
\EndOfBibitem
\bibitem[Kumar et~al.(2008)Kumar, Franzese, and {Eugene Stanley}]{Kumar2008a}
Kumar,~P.; Franzese,~G.; {Eugene Stanley},~H. \emph{J. Phys. Condens. Matter}
  \textbf{2008}, \emph{20}, 244114\relax
\mciteBstWouldAddEndPuncttrue
\mciteSetBstMidEndSepPunct{\mcitedefaultmidpunct}
{\mcitedefaultendpunct}{\mcitedefaultseppunct}\relax
\EndOfBibitem
\bibitem[Limmer and Chandler(2011)Limmer, and Chandler]{Limmer2011}
Limmer,~D.; Chandler,~D. \emph{J. Chem. Phys.} \textbf{2011}, \emph{135},
  134503\relax
\mciteBstWouldAddEndPuncttrue
\mciteSetBstMidEndSepPunct{\mcitedefaultmidpunct}
{\mcitedefaultendpunct}{\mcitedefaultseppunct}\relax
\EndOfBibitem
\end{mcitethebibliography}

\providecommand*\mcitethebibliography{\thebibliography}
\csname @ifundefined\endcsname{endmcitethebibliography}
  {\let\endmcitethebibliography\endthebibliography}{}

\end{document}